\documentclass[12pt,letterpaper]{article}
\pdfoutput=1

\usepackage{xcolor}
\usepackage{amsfonts,amssymb,amsmath}
\usepackage{MnSymbol,wasysym}
\usepackage{esint}
\usepackage{comment}

\usepackage{graphicx}
\usepackage{amsmath,systeme}
\usepackage{caption}
\usepackage{subcaption}
\usepackage{booktabs}

\usepackage{listings}

\usepackage{hyperref}

\usepackage[style=numeric-comp, sorting=none]{biblatex}
\DeclareFieldFormat{doi}{}
\DeclareFieldFormat{url}{}
\DeclareFieldFormat{eprint}{}

\numberwithin{equation}{section}

\newcommand{\be}{\begin{equation}}
\newcommand{\ee}{\end{equation}}
\newcommand{\ben}{\begin{displaymath}}
\newcommand{\een}{\end{displaymath}}
\newcommand{\bea}{\begin{eqnarray}}
\newcommand{\eea}{\end{eqnarray}}
\newcommand{\bean}{\begin{eqnarray*}}
\newcommand{\eean}{\end{eqnarray*}}

\usepackage{amsmath}

\newcommand{\beq}{\begin{equation}}
\newcommand{\eeq}{\end{equation}}
\newcommand{\beqr}{\begin{displaymath}}
\newcommand{\eeqr}{\end{displaymath}}
\newcommand{\beqa}{\begin{eqnarray}}
\newcommand{\eeqa}{\end{eqnarray}}
\newcommand{\beqar}{\begin{eqnarray*}}
\newcommand{\eeqar}{\end{eqnarray*}}

\newcommand{\cmmnt}[1]{}

\usepackage{setspace}

\usepackage{float}

\allowdisplaybreaks

\newcounter{customfootnote}  
\newcommand{\customfootnote}[1]
{%
  \refstepcounter{customfootnote}%
  \footnote[\value{customfootnote}]{#1}%
}

\begin{document}

\title{\LARGE \bf Self-Similar acoustic white hole solutions in Bose-Einstein condensates and their Borel analysis}

\author{
	 Sachin Vaidya$^1$\thanks{E-mail: \texttt{vaidya2@purdue.edu}} \\
	$^1$ Dep. of Physics and Astronomy, \\ Purdue University, W. Lafayette, IN  \\
	}

\date{}

\maketitle

\begin{abstract}
In this article, we study Self-Similar configurations of non-relativistic Bose-Einstein condensate (BEC) described by the Gross-Pitaevskii Equation (GPE). To be precise, we discuss singular Self-similar solutions of the Gross-Pitaevskii equation in 2D (with circular symmetry) and 3D (with spherical symmetry). We use these solutions to check for the crossover between the local speed of sound in the condensate and the magnitude of the flow velocity of the condensate, indicating the existence of a supersonic region and thus a sonic analog of a black/white hole. This is because phonons cannot go against the condensate flow from the supersonic to the subsonic region in such a system. We also discuss numerical techniques used and study the semi-analytical Laplace-Borel resummation of asymptotic series solutions while making use of the asymptotic transseries to justify the choice of numerical and semi-analytical approaches taken.
\end{abstract}

\clearpage
\newpage

\tableofcontents


\onehalfspacing
\section{Introduction}
\label{intro}
In \cite{PhysRevLett.46.1351} Unruh proposed studying Bose-Einstein condensates for gravitational analogs, which can be used to study phenomena such as Hawking radiation \cite{HawkingS.W.1975Pcbb}. Such Bose-Einstein condensate models have been studied \cite{Barceló2011, CarlosBarceló_2001, Visser2002, PhysRevLett.85.4643, PhysRevD.105.124066, Tian2022} in the past. Such models have also been developed in the cosmological context\customfootnote{Relativistic Bose-Einstein condensates can also be studied \cite{Dymnikova2001}, \cite{doi:10.1142/S0217732300002966}} (see \cite{PhysRevA.69.033602}, \cite{PhysRevLett.91.240407}) in the past. In this paper, we focus on developing a self-similar (scale invariant) analog gravity model from the Gross-Pitaevskii equation for non-relativistic BEC. 

Self-Similarity (Scale Invariance) is studied in the context of various systems in Physics. Self-Similarity of the Gross-Pitaevskii equation (GPE) was studied earlier in \cite{Barna2018}. Here we study a different type of self-similar configuration (with circular symmetry in 2D and spherical symmetry in 3D) in the Gross-Pitaevskii equation (GPE) that exhibits self-similarity in a combination of radial coordinate and time. 

To begin, we consider the Gross-Pitaevskii equation (GPE) with coupling $g$ and without an external potential (see ref. \cite{NozieresPinesBEC}), describing the dynamics of a Bose-Einstein condensate (BEC) 
\begin{equation} \label{e:GPE} 
i \hbar  \frac{\partial \psi  \left(\Vec{\mathbf{r}} , t\right)}{\partial t} = -\frac{\hbar^{2}}{2 m} 
\nabla^{2}\psi  \left(\Vec{\mathbf{r}} , t\right) + g {\lvert \psi  \left(\Vec{\mathbf{r}} , t\right)\rvert}^{2} \psi  \left(\Vec{\mathbf{r}} , t\right)
\end{equation}

Small perturbations in the condensate can be shown to give an approximate local speed of sound in the long wavelength limit as (see ref. \cite{NozieresPinesSound})
\begin{equation} \label{e:speed of sound variable}
c\left(\Vec{\mathbf{r}} , t\right)=\sqrt{\frac{n\left(\Vec{\mathbf{r}} , t\right)g}{m}}
\end{equation}
and similarly from the current density defined using the Gross-Pitaevskii equation, we can obtain the flow velocity (see ref. \cite{NozieresPinesVelocity}) as follows
\begin{equation} \label{e:flow velocity}
\Vec{v} \left(\Vec{\mathbf{r}} , t \right) = \frac{\Vec{\mathbf{j}} \left(\Vec{\mathbf{r}} , t\right)}{n \left(\Vec{\mathbf{r}} , t\right)} = \frac{\hbar}{m} \Vec{\nabla} \theta  \left(\Vec{\mathbf{r}} , t\right)
\end{equation}
We use \eqref{e:GPE}, \eqref{e:speed of sound variable}, and \eqref{e:flow velocity} for a reference for the rest of the calculations. We focus on finding self-similar (in $r$ and $t$) solutions to the Gross-Pitaevskii equation \eqref{e:GPE} to look for solutions that can correspond to an acoustic black/white hole. It turns out that we cannot solve this problem analytically. Therefore, we study (in 2D) the perturbative series solution, which turns out to be factorially divergent, as we shall see. We use this series to demonstrate the utility of Laplace-Borel resummation to get an approximate self-similar solution from limited asymptotic data. Furthermore, we will also see that this series turns out to have non-perturbative corrections resulting in a transseries (see Ref. \cite{10.1155/S1073792895000286}), the properties of which we briefly discuss. We use this transseries to justify the use of the Runge-Kutta method of order four (RK4) \cite{Hairer2015} for numerical solutions.

\section{Self similar solutions}
To study the Self-Similar solutions with radial and time dependence, we write the Gross-Pitaevskii Equation (GPE) \eqref{e:GPE} in the form of scaled variables as 
\begin{equation} \label{e:SelfSimilarPDE}
i \frac{\partial \chi  \left(R , T\right)}{\partial T}
 = 
-\nabla_R^{2}\chi  \left(R , T\right)+{| \chi  \left(R , T\right)|}^{2} \chi  \left(R , T\right)
\end{equation}
where the scaled variables are 
\beqa
T &=& t, \\ 
R&=&\sqrt{\frac{2m}{\hbar}}r, \\
\chi  \left(R , T\right) &=& \sqrt{\frac{g}{\hbar}}\ \psi  \left(R , T\right), \\
\rho  \left(R , T\right) &=& \sqrt{\frac{g}{  \hbar}}\ \sqrt{n  \left(R , T\right)} = \sqrt{\frac{g}{  \hbar}}\ \lvert \psi  \left(R , T\right)\rvert
\eeqa

It can be shown that the solution of the form $\chi  \left(R , T\right) = T^{\alpha} \zeta  \left(R^{\beta} T^{\gamma}\right)$ (similar to \cite{Barna2018}) exists if $\alpha=-\frac{1}{2}$ and $\gamma=-\frac{\beta}{2}$. Changing the coordinates to $v=\frac{R}{\sqrt{T}}$ and substituting 
\beq
\chi  \left(R , T\right) = \frac{1}{\sqrt{T}} \zeta  \left(v \right) = \frac{1}{\sqrt{T}} \eta  (v) \exp(i \theta(v)))
\eeq
into~\eqref{e:SelfSimilarPDE}, we get the imaginary part.
\begin{equation} \label{e:SelfSimilarImODE}
\left(2 \frac{d \eta  \left(v \right)}{d v}+\frac{\eta  \left(v \right) (\mathbf{d}-1)}{v}\right) \left(\frac{d \theta  \left(v \right)}{d v}\right)-\frac{v}{2}\frac{d \eta  \left(v \right)}{d v}+\eta  \left(v \right) \left(\frac{d^{2} \theta  \left(v \right)}{d v^{2}}\right)-\frac{\eta  \left(v \right)}{2} = 0 
\end{equation}
and the real part
\begin{equation} \label{e:SelfSimilarReODE}
\frac{\mathbf{d}-1}{v} \left(\frac{d \eta  \left(v \right)}{d v}\right)-\eta  \left(v \right)^{3}+\frac{d^{2} \eta  \left(v \right)}{d v^{2}}-\eta  \left(v \right) \left(\frac{d \theta  \left(v \right)}{d v}\right)^{2}+\frac{v \eta  \left(v \right)}{2} \left(\frac{d \theta  \left(v \right)}{d v}\right) = 0
\end{equation}
Although in $\mathbf{d}=3$~\eqref{e:SelfSimilarImODE} does not have a closed-form solution, in $\mathbf{d}=2$ it does have a closed-form solution as follows
\beq \label{e:dtheta_dv}
\frac{d \theta  \left(v \right)}{d v} = \frac{v}{4} + \frac{B}{v \eta  \left(v \right)^{2}}
\eeq

Furthermore, in the new scaled variables, the scaled local speed of sound (from \eqref{e:speed of sound variable}) becomes
\beq \label{e:scaled_sound_ss}
C \left(R, T\right) = \sqrt{2} \rho  \left(R, T\right) = \sqrt{2} \lvert \chi  \left(R, T\right) \rvert = \sqrt{2} \frac{1}{\sqrt{T}} \eta  \left(\frac{R}{\sqrt{T}} \right) = \sqrt{2} \frac{1}{\sqrt{T}} \eta  \left(v \right)
\eeq
and the scaled radial flow velocity (from \eqref{e:flow velocity}) of Bose-Einstein condensate (BEC) becomes
\beq \label{e:scaled_vel_ss}
\Vec{\mathbf{V}}  \left(R, T \right) = 2\frac{\partial \theta  \left(R, T \right)}{\partial R} \hat{\mathbf{R}} = 2\frac{\partial}{\partial R}  \left(\theta \left(\frac{R}{\sqrt{T}} \right) \right) \hat{\mathbf{R}} = 2\frac{1}{\sqrt{T}}\frac{d \left(\theta \left(v \right) \right)}{d v} \hat{\mathbf{R}}
\eeq

\section{Perturbative series Analysis} 
\begin{figure}[H]
    \centering
    \begin{subfigure}{1\textwidth}
        \centering
        \includegraphics[width=\linewidth]{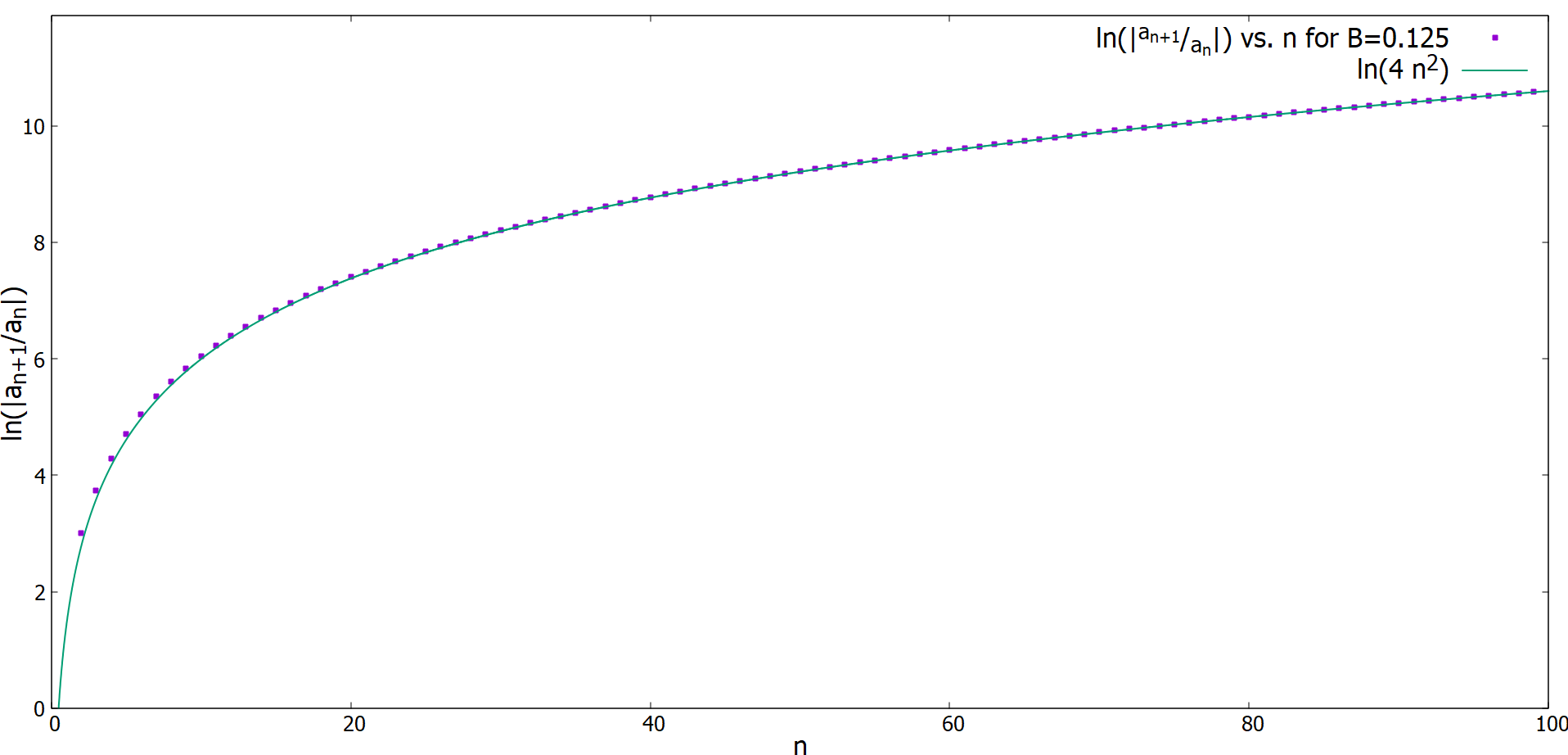}
        \caption{$B = 0.125$} \label{fig:B0.125_coefficient_growth_SelfSimilar}
    \end{subfigure}
    \\[10pt]
    \begin{subfigure}{1\textwidth}
        \centering
        \includegraphics[width=\linewidth]{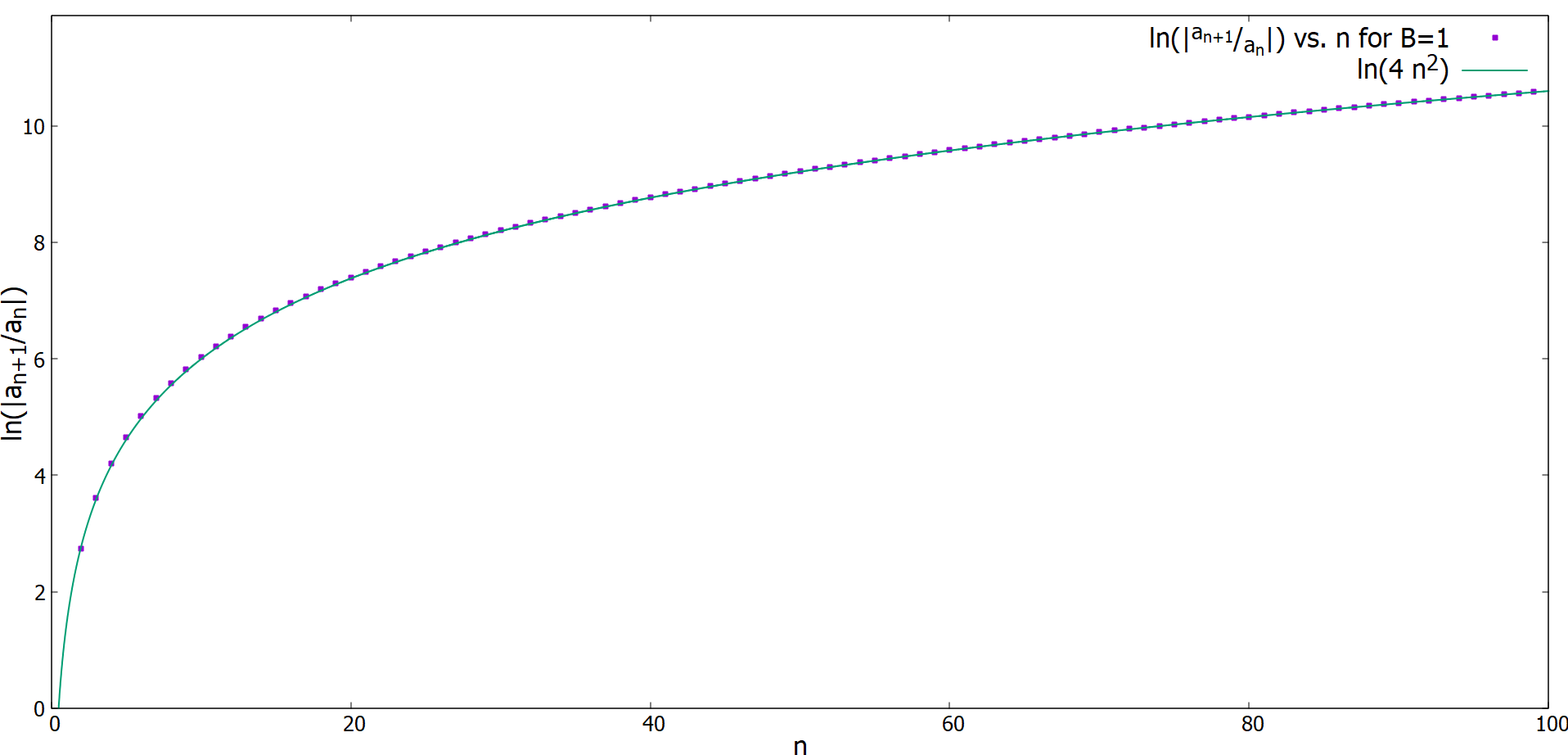}
        \caption{$B = 1$} \label{fig:B1_coefficient_growth_SelfSimilar}
    \end{subfigure}
    \vspace{5pt}
    \caption{Sample plots of $\ln\left(\left|\frac{a_{n+1}}{a_n}\right|\right)$ vs. $n$ for Self Similar solutions in 2D}
    \label{fig:coefficient_growth_SelfSimilar}
\end{figure}

For $\mathbf{d}=2$, equations~\eqref{e:SelfSimilarReODE} and \eqref{e:dtheta_dv} result in
\begin{equation} \label{e:2DSelfSimilarODE}
\frac{d^{2} \eta  \left(v \right)}{d v^{2}}+\frac{1}{v} \frac{d \eta  \left(v \right)}{d v}+\frac{\eta  \left(v \right) v^{2}}{16}-\frac{B^{2}}{\eta  \left(v \right)^{3} v^{2}}-\eta  \left(v \right)^{3} = 0
\end{equation}
The asymptotic perturbative series solution of~\eqref{e:2DSelfSimilarODE} can be written as follows.
\begin{align}\label{e:2DSelfSimilarODE_series_v}
\lefteqn{\eta \left(v \right) = \frac{2 \sqrt{B}}{v} \left(1 + \left(B -\frac{1}{4}\right) \left(\frac{16}{v^{4}}+\left(\frac{9 B}{2}-\frac{53}{8}\right) \frac{256}{v^{8}}\right.\right.}\nonumber\\&\indent\indent\indent\indent\indent\indent\left.\left.+\left(\frac{55}{2} B^{2}-\frac{535}{4} B +\frac{4447}{32}\right) \frac{4096}{v^{12}}+\ldots \right)\right)
\end{align}
Since the solution must be real positive, $B$ must be real positive to ensure that $\sqrt{B}$ is real positive. It can be shown that the series coefficients have an alternating sign at large order. Also, as can be seen in Figures \ref{fig:B0.125_coefficient_growth_SelfSimilar} and \ref{fig:B1_coefficient_growth_SelfSimilar}, at large order $\left|\frac{a_{n+1}}{a_n}\right| \approx \frac{(2(n+1))!}{(2n)!} \to 4 n^2$ which means the coefficients grow like $\approx\left(-1\right)^n(2n)!$ at large order up to a multiplying constant. 

Note that the series lacks an arbitrary integration constant similar to the stationary singular solutions discussed in \cite{vaidya2024stationaryacousticblackhole}. Asymptotically, the leading term in the solution goes like $\frac{2\sqrt{B}}{v}$ as $v\to\infty$ and therefore substituting $\eta  \left(v \right) = \frac{2\sqrt{B}(1 + A f  \left(v \right))}{v}$ into \eqref{e:2DSelfSimilarODE} and linearizing it in $f(v)$ (see \cite{Costin2015}), we get $f \left(v \right) \approx \exp(\frac{\pm i v^{2}}{4}) + \ldots$ where $A$ represents two missing arbitrary constants of integration corresponding to $\exp(\frac{+ i v^{2}}{4})$ and $\exp(\frac{- i v^{2}}{4})$ respectively. This reveals the arbitrary constants that were missing earlier. The functions $\exp(\frac{\pm i v^{2}}{4})$ are non-perturbative as $v\to\infty$ and they are finite oscillatory everywhere on the real $v$ line meaning that they don't become singular. Therefore, none of the two arbitrary constants they bring in with them, has to be $0$. Furthermore, since the differential equation is nonlinear in the dependent variable, non-perturbative corrections appear with ever-increasing powers and also bring in their own factorially divergent perturbative series multiplying them. This is how we end up with a two-parameter transseries (similar to \cite{Aniceto2015} and \cite{10.1093/imrn/rnr029}) on the real $v$ line in this system. In addition to that, since the self-similar perturbative series has two types of non-perturbative corrections, there also turn out to be terms containing multiplications of these two types of non-perturbative corrections (instanton anti-instanton interactions).

With $\eta \left(v \right)  = \frac{2 \sqrt{B}}{v} g \left(v \right)$ and a change of variable $u = \frac{4}{v^{2}}$ in~\eqref{e:2DSelfSimilarODE_series_v}, we get
\begin{equation} \label{e:2DSelfSimilarODE_series}
g \left(u \right) = 1 + \left(B -\frac{1}{4}\right) \left(u^{2}+\left(\frac{9 B}{2}-\frac{53}{8}\right) u^{4}+\left(\frac{55}{2} B^{2}-\frac{535}{4} B +\frac{4447}{32}\right) u^{6}\right)+\ldots
\end{equation}
and two-parameter transseries for this system (including \eqref{e:2DSelfSimilarODE_series}) with a first few terms turns out to be as follows:
\begin{align}\label{e:2DSelfSimilarODE_transseries_u}
g \left(u \right) & = 1 + \left(B -\frac{1}{4}\right) \left(u^{2}+\left(\frac{9 B}{2}-\frac{53}{8}\right) u^{4}+\left(\frac{55}{2} B^{2}-\frac{535}{4} B +\frac{4447}{32}\right) u^{6}\right)+\ldots \nonumber \\
&+ C_{1} {\mathrm e}^{\frac{\mathrm{i}}{u}} \left(1+\frac{\mathrm{i} \left(6 B -1\right) u}{2}-\frac{3 \left(12 B^{2}-8 B +1\right) u^{2}}{8}+\ldots\right) \nonumber \\
&+ C_{2} {\mathrm e}^{\frac{\mathrm{-I}}{u}} \left(1-\frac{\mathrm{i} \left(6 B -1\right) u}{2}-\frac{3 \left(12 B^{2}-8 B +1\right) u^{2}}{8}+\ldots\right) \nonumber \\
&+ C_{1} C_{2} \left(3+\left(18 B -\frac{3}{4}\right) u^{2}+\left(159 B^{2}-144 B +\frac{159}{32}\right) u^{4}+\ldots\right) \nonumber \\
& +\ldots 
\end{align}
where $C_{1,2}$ are complex and $C_2=C_1^*$ to ensure that the transseries represents a real function\customfootnote{Only few terms are displayed in the transseries here. However, in principle we can easily get first few hundred term using a software package such as Maple.}. Thus we still have two real free parameters in the transseries. There are no other constraints on these two real free parameters for the transseries solutions to be real. And therefore, we can set both the real parameters to $0$ (effectively using \eqref{e:2DSelfSimilarODE_series}) to demonstrate the Laplace-Borel resummation in 2D without violating the reality condition on the solution, as shown in the subsequent section. On the other hand, setting these two real parameters to have any arbitrary nonzero values would result in oscillatory self-similar solutions with the oscillations of the form $\exp(\frac{\pm i}{u}) = \exp(\frac{\pm i v^{2}}{4})$ as $v\to\infty$ as we shall see. Furthermore, the freedom to choose two real free parameters (with zero or nonzero arbitrary values) for a second order ODE is equivalent to picking a function and its slope at any large $v$ and solving the problem as an initial value problem using methods such as RK4 \cite{Hairer2015}. Therefore, numerically we solve this problem as an IVP using RK4 in C++.

In $\mathbf{d}=3$, there is no closed-form solution for $\frac{d \theta  \left(v \right)}{d v}$. However, in principle, a similar procedure as $\mathbf{d} = 2$ can still be adopted. Here we make the simplest estimates as $v\to\infty$ for the purposes of the numerical calculations.
\begin{align} \label{e:asymptotic3D}
\text{When }\ \ \eta  (v)  \approx \frac{C}{v} &\implies \frac{d \theta  \left(v \right)}{d v} \approx \frac{2 C^2}{v^3} \nonumber \\
\text{and when }\ \ \eta  (v) \approx \frac{C}{v^2} &\implies \frac{d \theta  \left(v \right)}{d v} \approx \frac{v}{2} \nonumber \\
\end{align}
where $C>0$ to ensure $\eta  (v) > 0$. We use these conditions at large $v$ as a reference to solve the initial value problem for a system of coupled ODEs in 3D using RK4.

\begin{figure}[H]
\centering
    \includegraphics[width=\linewidth]{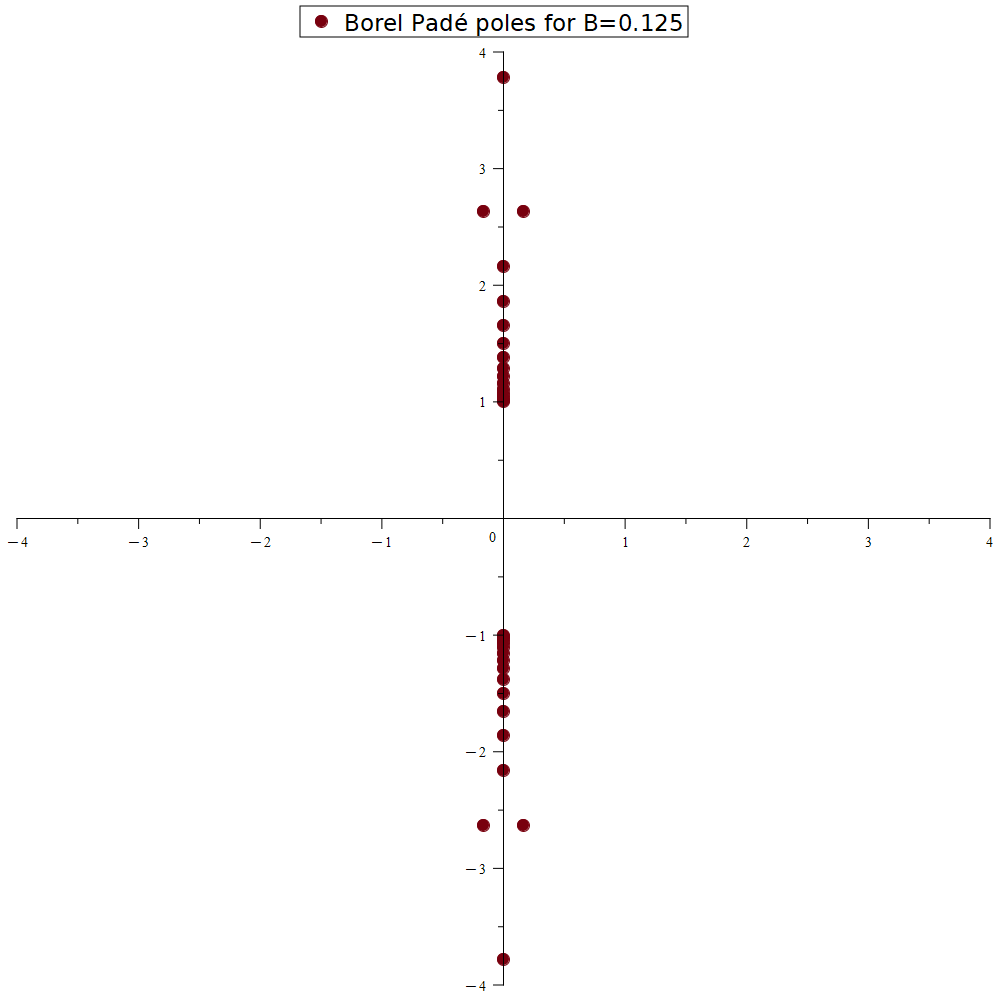}
    \vspace{5pt}
    \caption{Approximate Borel Pad\'e poles of Borel transform of \eqref{e:2DSelfSimilarODE_series} in $U$ coordinate}\label{fig:Borel Pade poles Self Similar}
\end{figure}

\subsection{Laplace-Borel Analysis}
\begin{figure}[H]
    \centering
    \begin{subfigure}{1\textwidth}
        \centering
        \includegraphics[width=\linewidth]{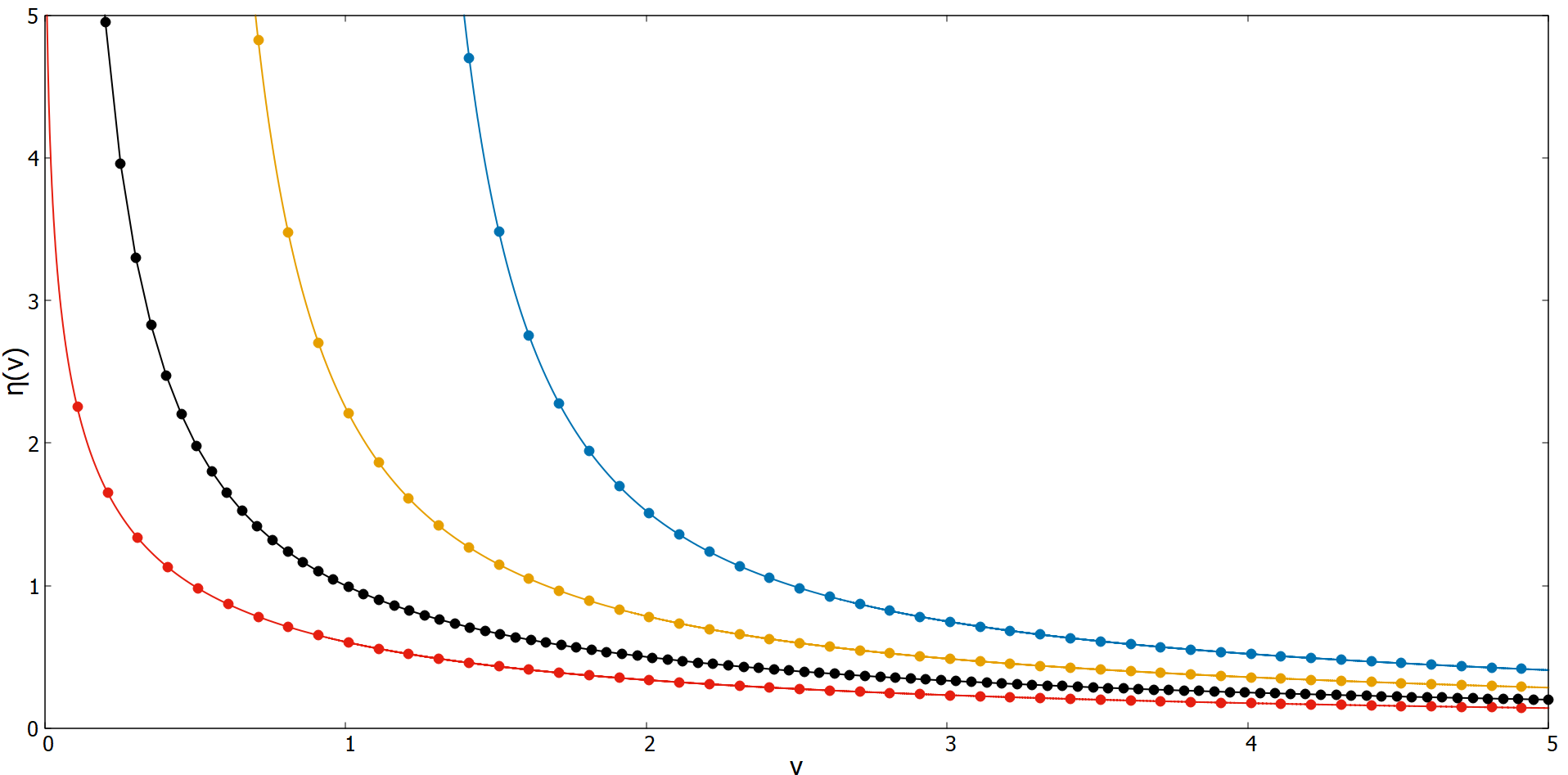}
        \caption{$\eta(v)$ vs. $v$} \label{fig:rho_v}
    \end{subfigure}
    \\[10pt]
    \begin{subfigure}{1\textwidth}
        \centering
        \includegraphics[width=\linewidth]{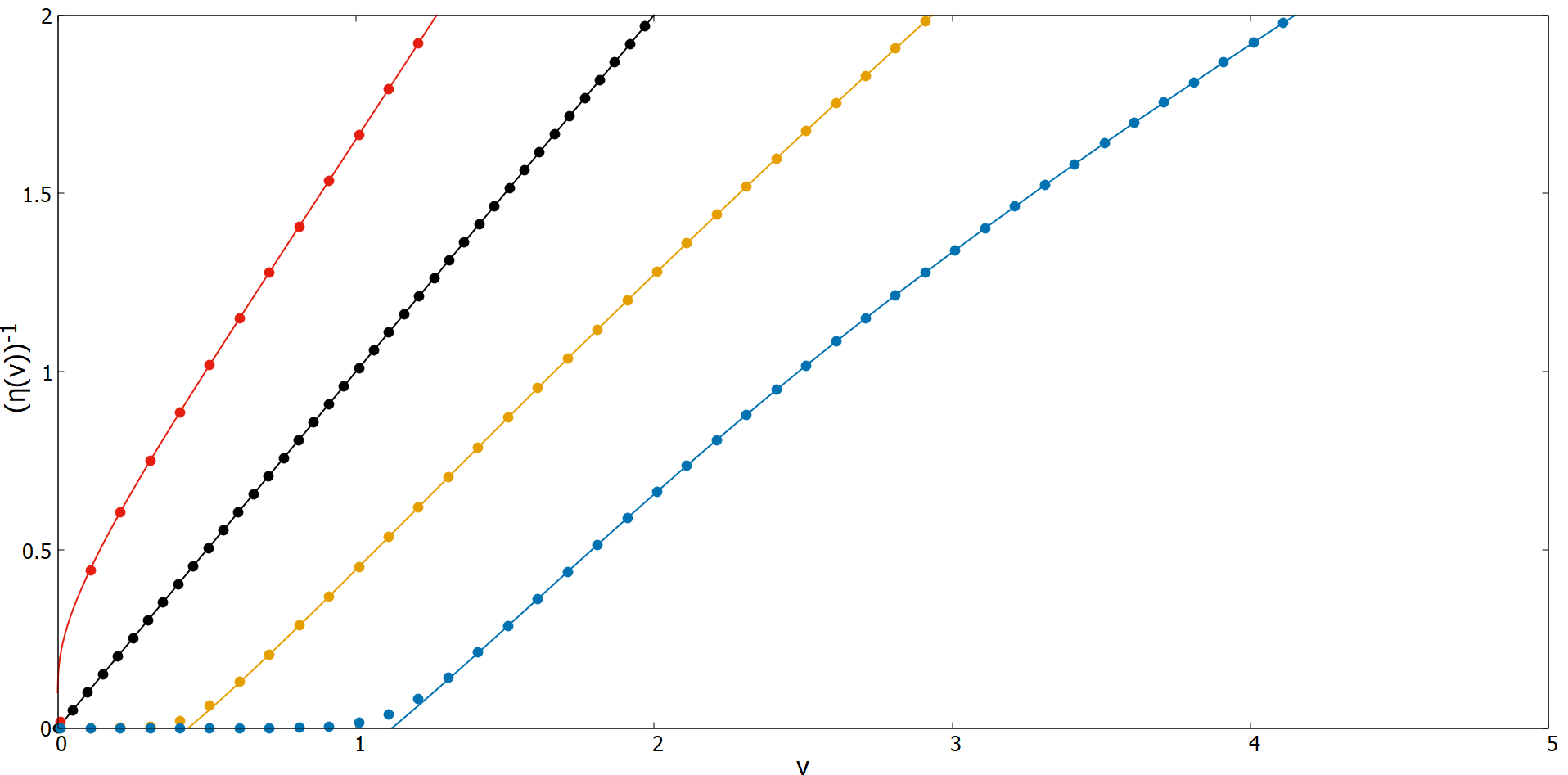}
        \caption{$\frac{1}{\eta(v)}$ vs. $v$} \label{fig:Inv_rho_v}
    \end{subfigure}
    \vspace{5pt}
    \caption{Initial value solution matched with Conformal Laplace-Borel transform for various values of parameter B in 2D. From left to right\protect\hyperlink{fn:6}{\protect\footnotemark[6]} $B=0.125$,$0.25$,$0.5$,$1.0$. The circles depict the resummed asymptotic series solution and the solid lines are the numerical solutions. In plot (b) we can see the discrepancy between the two for the same solutions when $\rho\rightarrow\infty$ (near the horizontal axis where the series resummation fails).} \label{fig:2D_SelfSimilar_resummation}
\end{figure}

We evaluate the Borel transform\customfootnote{$U$ is the Borel variable and we consider $b=100$ for these calculations} of $\frac{g(u)}{\left(B-\frac{1}{4}\right)}$ of order $\mathcal{O}(U^{2b})$ from~\eqref{e:2DSelfSimilarODE_series} and convert it into the Pad\'e approximation or the Conformal Pad\'e approximation using a procedure similar to Stationary solutions in \cite{vaidya2024stationaryacousticblackhole}. Since the series in~\eqref{e:2DSelfSimilarODE_series}, at large order, has factorially growing coefficients (see Fig. \ref{fig:coefficient_growth_SelfSimilar}) with alternating sign, there is no singularity on the positive real $U$ line in the Borel plane (similar to the Painlev\'e equation in \cite{Costin_2019}) as can be seen from the poles of the Borel Pad\'e approximation of \eqref{e:2DSelfSimilarODE_series} in figure~\ref{fig:Borel Pade poles Self Similar}. Thus, the Laplace transform after the Borel Pad\'e or the conformal Borel Pad\'e transform does not have imaginary ambiguities containing real exponential functions (see \cite{vaidya2024stationaryacousticblackhole} for reference). In addition to that, the non-perturbative corrections $\left(\exp(\frac{\pm i}{u}) = \exp(\frac{\pm i v^{2}}{4})\right)$ to the asymptotic perturbative series do not contain any real exponential functions and are finite. This indicates consistency. Therefore, no cancellation of imaginary non-perturbative ambiguities containing real exponents is needed for the solution to be real. Thus, for the purpose of demonstrating the Laplace-Borel resummation (see Fig. \ref{fig:2D_SelfSimilar_resummation}), we can restrict ourselves to solutions with both parameters of the transseries being $0$, without violating the reality condition on $\eta  \left(v \right)$. 

\begin{figure}
    \centering
    \begin{subfigure}{0.8\textwidth}
        \centering
        \includegraphics[width=\linewidth]{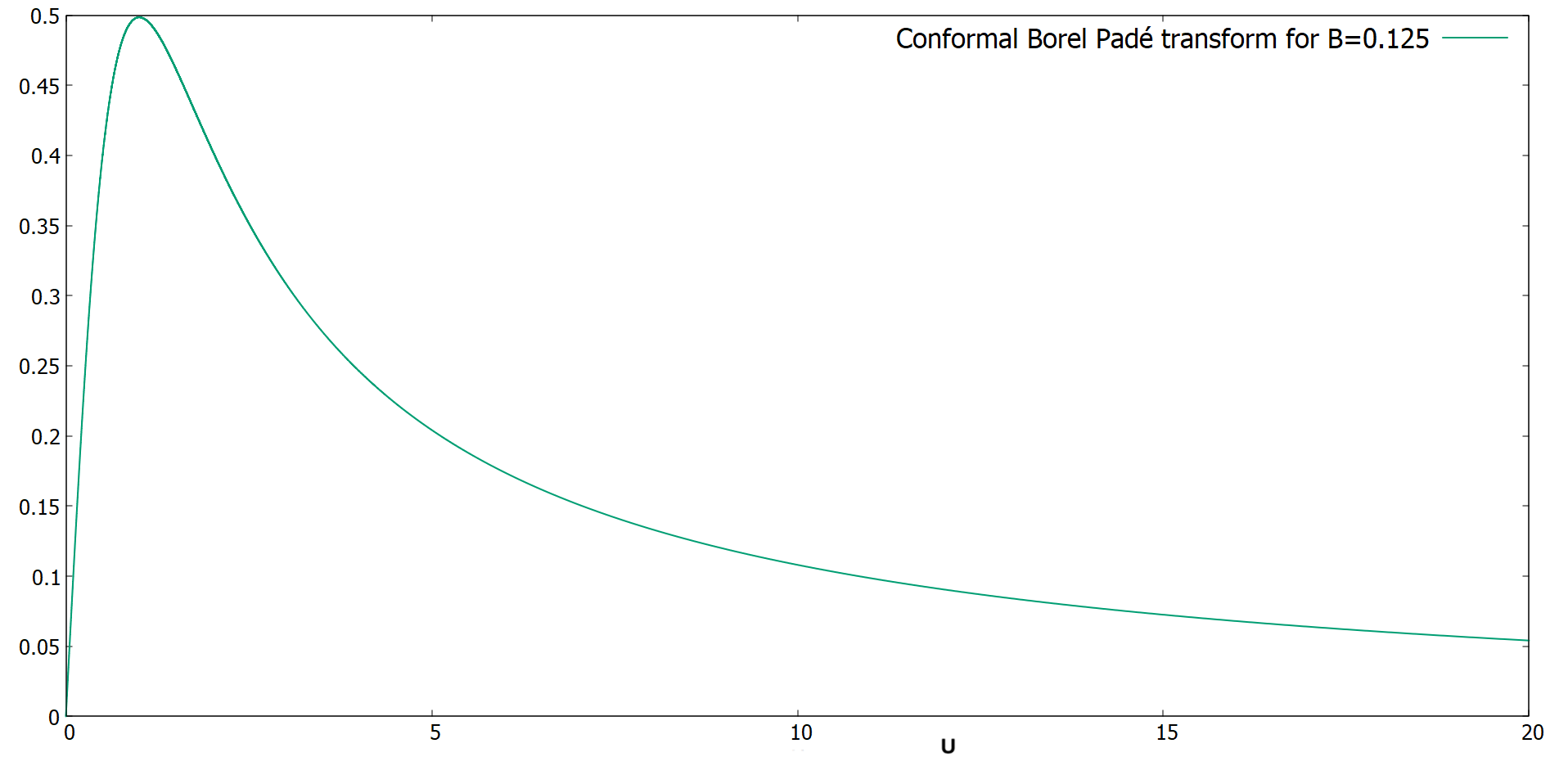}
        \caption{Conformal Borel Pad\'e transform for $B=0.125$} \label{fig:B_0.125_CBP}
    \end{subfigure}
    \\[8pt]
    \begin{subfigure}{0.8\textwidth}
        \centering
        \includegraphics[width=\linewidth]{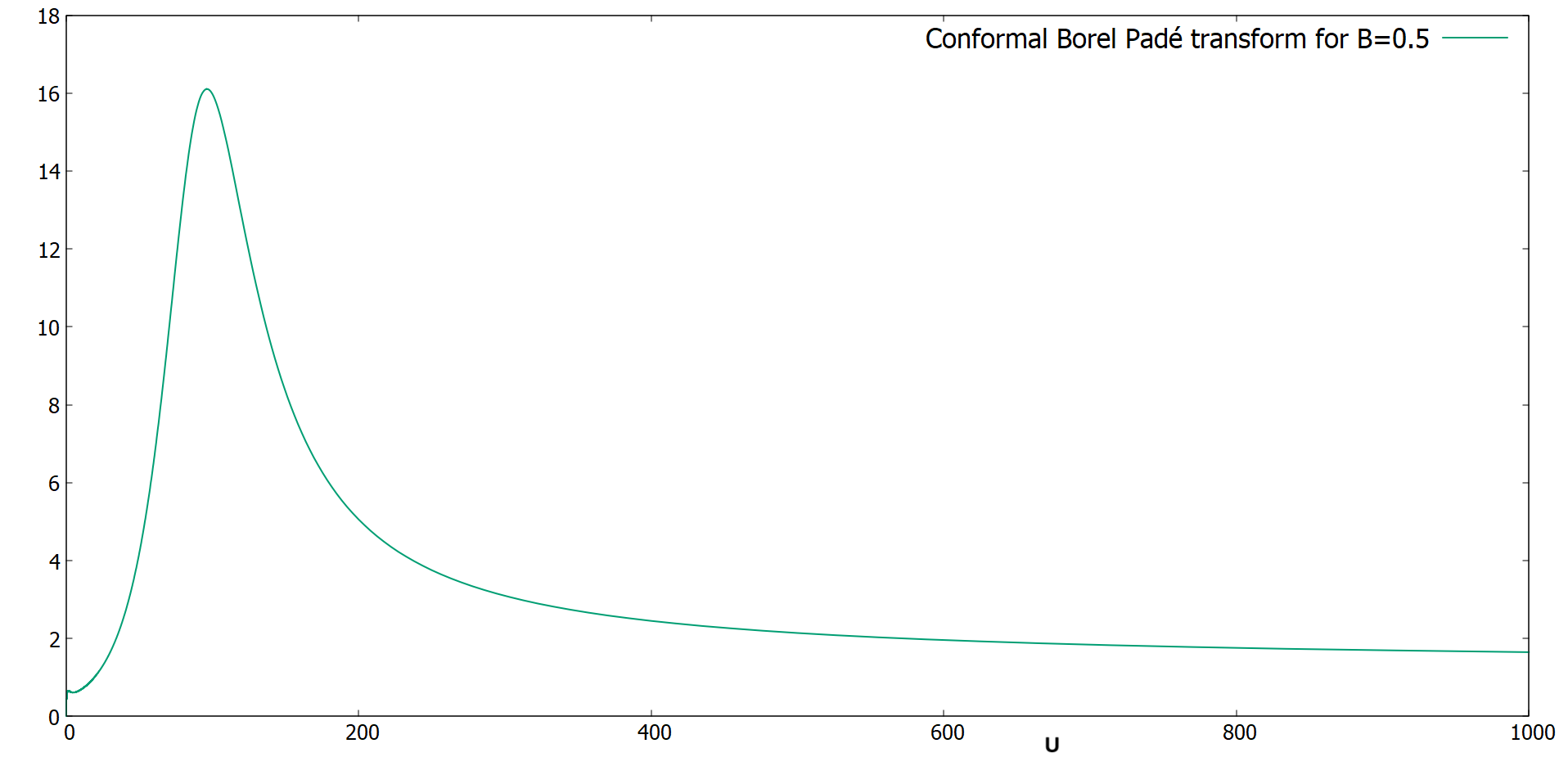}
        \caption{Conformal Borel Pad\'e transform for $B=0.5$} \label{fig:B_0.5_CBP}
    \end{subfigure}
	\\[8pt]
    \begin{subfigure}{0.8\textwidth}
        \centering
        \includegraphics[width=\linewidth]{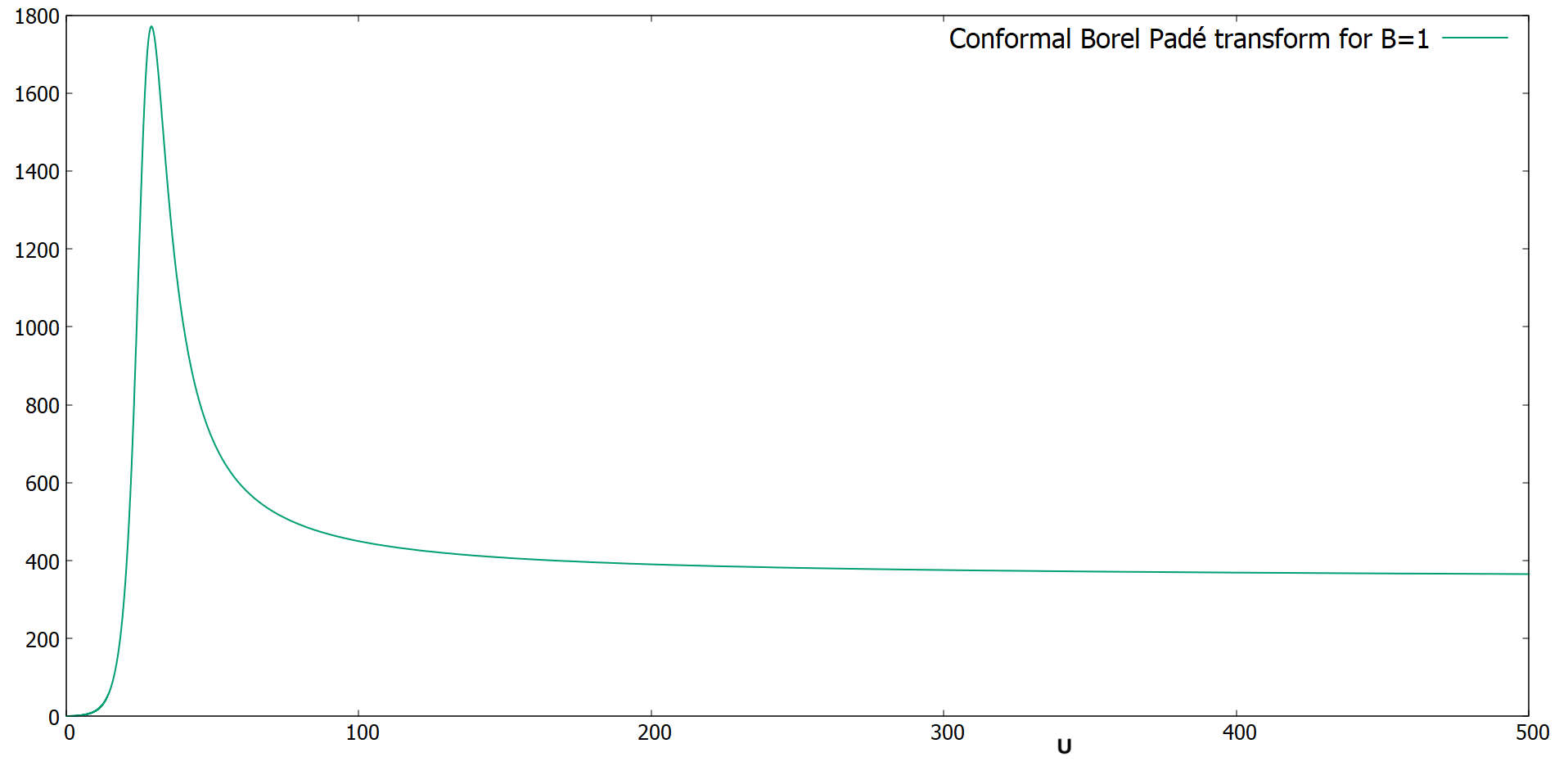}
        \caption{Conformal Borel Pad\'e transform for $B=1$} \label{fig:B_1_CBP}
    \end{subfigure}
    \vspace{3pt}
    \caption{Conformal Borel Pad\'e transforms for $B=0.125$, $0.5$, $1$ respectively in 2D} \label{fig:Conformal Borel Pade transforms}
\end{figure}

For the actual calculations, similar to \cite{vaidya2024stationaryacousticblackhole}, we use conformal\customfootnote{With conformal map $U=\frac{2 z}{1-z^2}$ similar to \cite{Costin_2019}} Borel-Pad\'e approximation\customfootnote{$(1+z^2)$ because the $g(u)$ series has even powers of $u$ and the leading Borel singularities in variable $z$ are at $U=z=\pm i$} of $(1+z^2) G(z)$ of order $[b+1,b]$ where $G(z)$ is the conformally mapped Borel transform of $\frac{g(u)}{\left(B-\frac{1}{4}\right)}$ of order $\mathcal{O}(z^{2b})$. Then we divide this result by $(1+z^2)$ to get the actual conformal Borel-Pad\'e transform of order $[b+1,b+2]$ (see Fig. \ref{fig:Conformal Borel Pade transforms}) of $\frac{g(u)}{\left(B-\frac{1}{4}\right)}$, substitute $z$ in terms of $U$, and Laplace integrate it on the real positive $U$ line in the Borel plane. As is clearly evident from the conformal Borel-Pad\'e transforms shown in Fig. \ref{fig:Conformal Borel Pade transforms}, there is no singularity on the Laplace integration path in the Borel plane and therefore no singularity subtraction or contour deformation is needed, again indicating that there is no imaginary ambiguity in the Laplace integration result. Laplace integration is performed by integrating over $U \in [0,1]$ and over $x = \frac{1}{U} \in [0,1]$ to get the Laplace-Borel resummation result (see Fig. \ref{fig:2D_SelfSimilar_resummation}). Similarly, the Laplace-Borel resummation can also be done for $\mathbf{d} = 3$ in principle.

\clearpage
\hypertarget{fn:6}{\footnotetext[6]{For $B=0.25$, $g(u)=1$ and therefore no Laplace Borel resummation is needed}}

\section{Numerical Solution}
\begin{figure}[H]
   \centering
    \includegraphics[width=\linewidth]{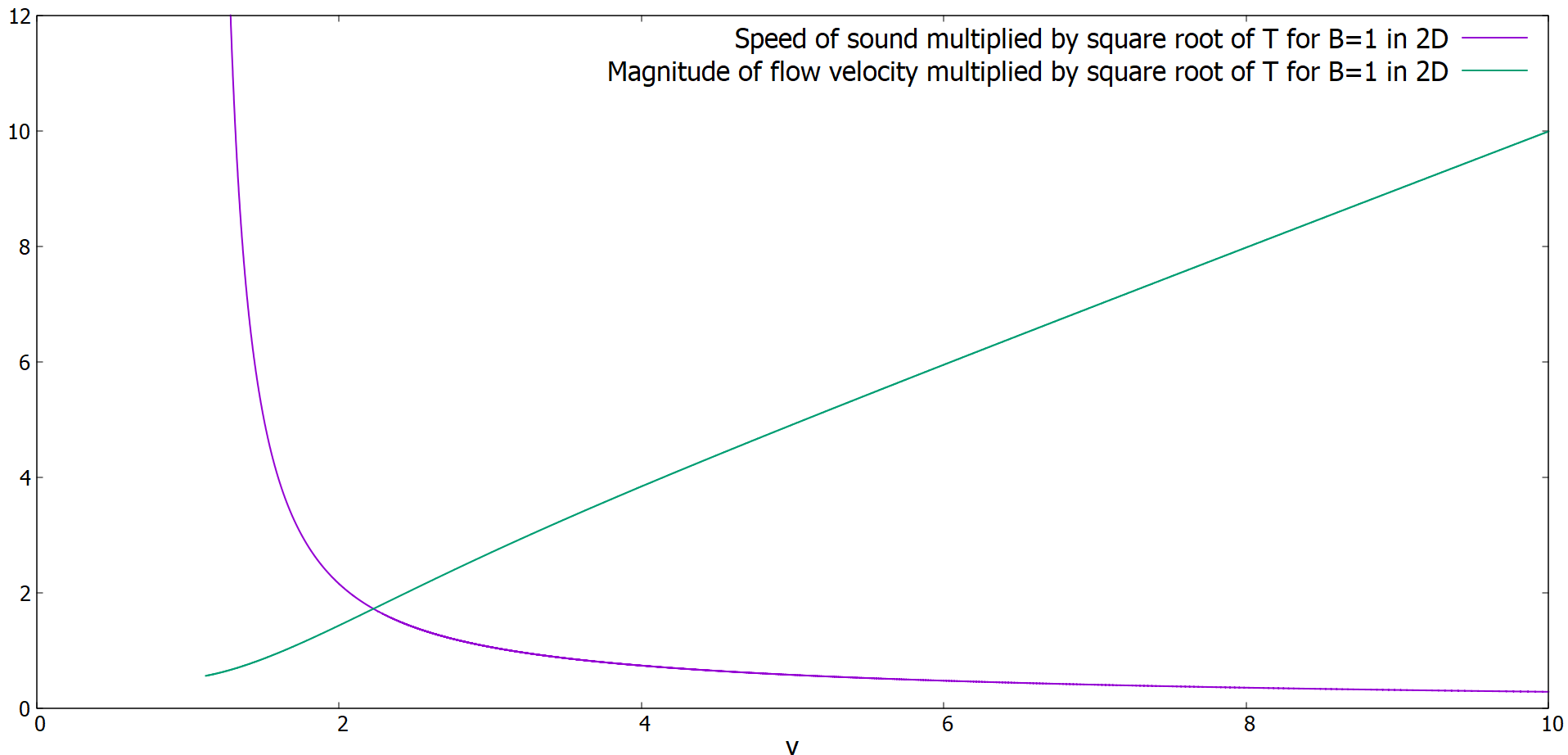}
    \vspace{-5pt}
    \caption{Scaled sound speed and flow speed of BEC in coordinate $v$ for $B = 1$ in 2D}
    \label{fig:2D_SelfSimilar_numerical}
\end{figure}

Similar to the stationary case in \cite{vaidya2024stationaryacousticblackhole}, it can be shown that the governing differential equation \eqref{e:2DSelfSimilarODE} in 2D can have singular solutions for $\eta(v)$. To be precise, in 2D the singularity is of the form $\eta  \left(v \right) \approx \frac{1}{v}$ when it occurs at the origin, which gives $\frac{d \theta  \left(v \right)}{d v}\Bigr|_{\substack{v=0}} = 0$. And (in 2D) the singularity is of the form $\eta  \left(v \right) \approx \frac{\sqrt{2}}{v-v_0}$ when it occurs at $v_0>0$, which gives $\frac{d \theta  \left(v \right)}{d v}\Bigr|_{\substack{v=v_{0}}} = \frac{v_0}{4}$. This is in agreement with the initial value solution for $\eta  \left(v \right)$ starting from $v=100$ with different values of $B$ (see Fig. \ref{fig:2D_SelfSimilar_resummation}) where the initial conditions are determined by the series \eqref{e:2DSelfSimilarODE_series_v}. Logarithmically singular solution at $v=0$ is also possible in 2D, as can be seen for $B=0.125$ in Figure~\ref{fig:2D_SelfSimilar_resummation}. For $B=\frac{1}{4}$, there exists an exact solution $\eta  \left(v \right) = \frac{1}{v}$ in 2D. We solve these ODEs as initial value problems starting from $v=100$ using RK4 \cite{Hairer2015} and initial conditions determined by the series \eqref{e:2DSelfSimilarODE_series_v} as shown for 2D in the Figure \ref{fig:2D_SelfSimilar_resummation} for example. In 2D, Laplace-Borel resummation agrees well with the numerical solutions shown in Figure \ref{fig:2D_SelfSimilar_resummation}. Scaled speed of sound and magnitude of flow velocity for a sample solution (in 2D) with $B = 1$ are shown in Figure \ref{fig:2D_SelfSimilar_numerical}. 

\begin{figure}[H]
    \centering
    \begin{subfigure}{1\textwidth}
        \centering
        \includegraphics[width=\linewidth]{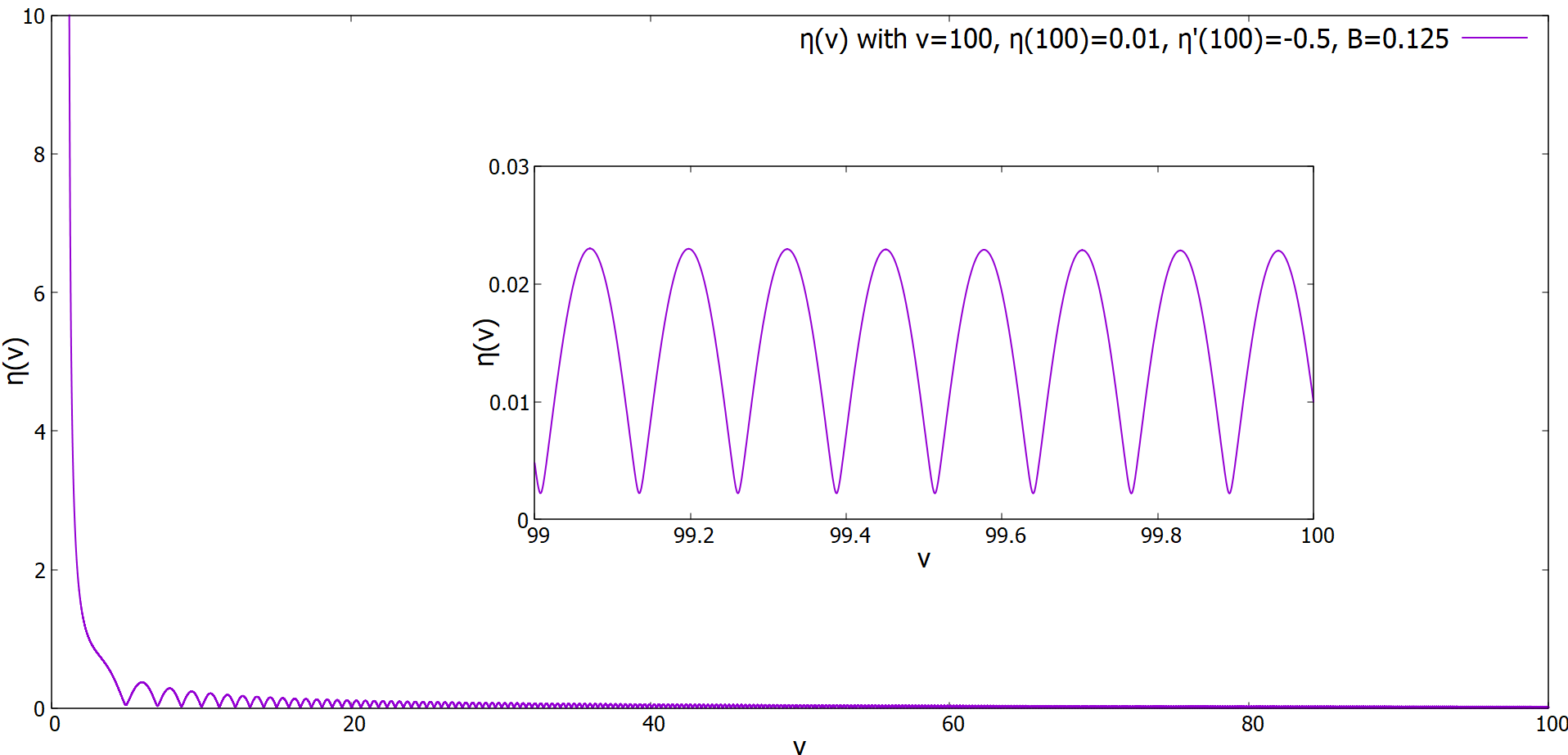}
        \caption{Sample solution} \label{fig:B_0.125_eta_0.01_etaPrime_-0.5}
    \end{subfigure}
    \\[10pt]
    \begin{subfigure}{1\textwidth}
        \centering
        \includegraphics[width=\linewidth]{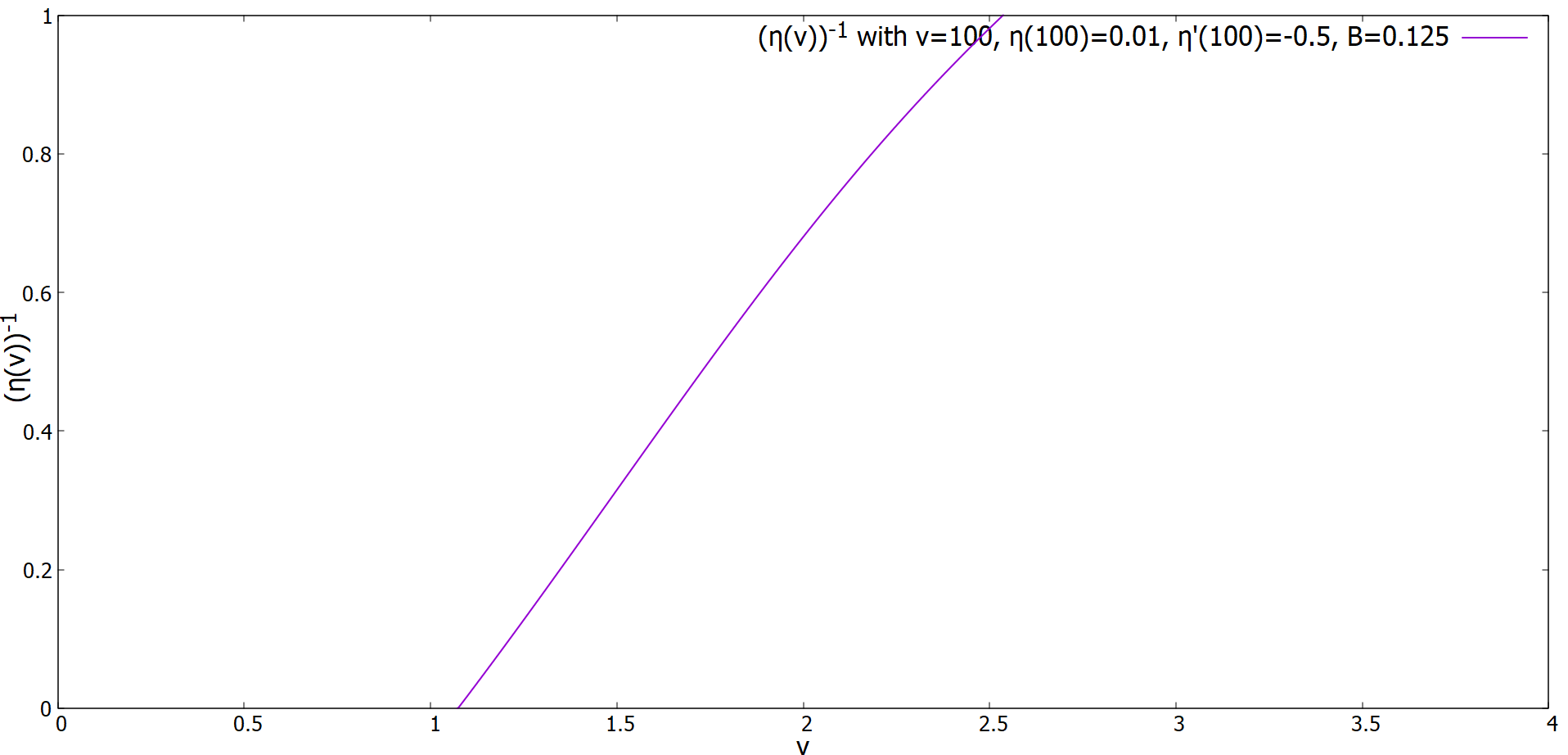}
        \caption{Inverted sample solution} \label{fig:B_0.125_eta_0.01_etaPrime_-0.5_etaInv}
    \end{subfigure}
    \vspace{5pt}
    \caption{Sample oscillatory solution with singularity of the form $\frac{\sqrt{2}}{v-v_{0}}$ in 2D}
    \label{fig:eta_0.01_etaPrime_-0.5}
\end{figure}

\begin{figure}[H]
    \centering
    \begin{subfigure}{1\textwidth}
        \centering
        \includegraphics[width=\linewidth]{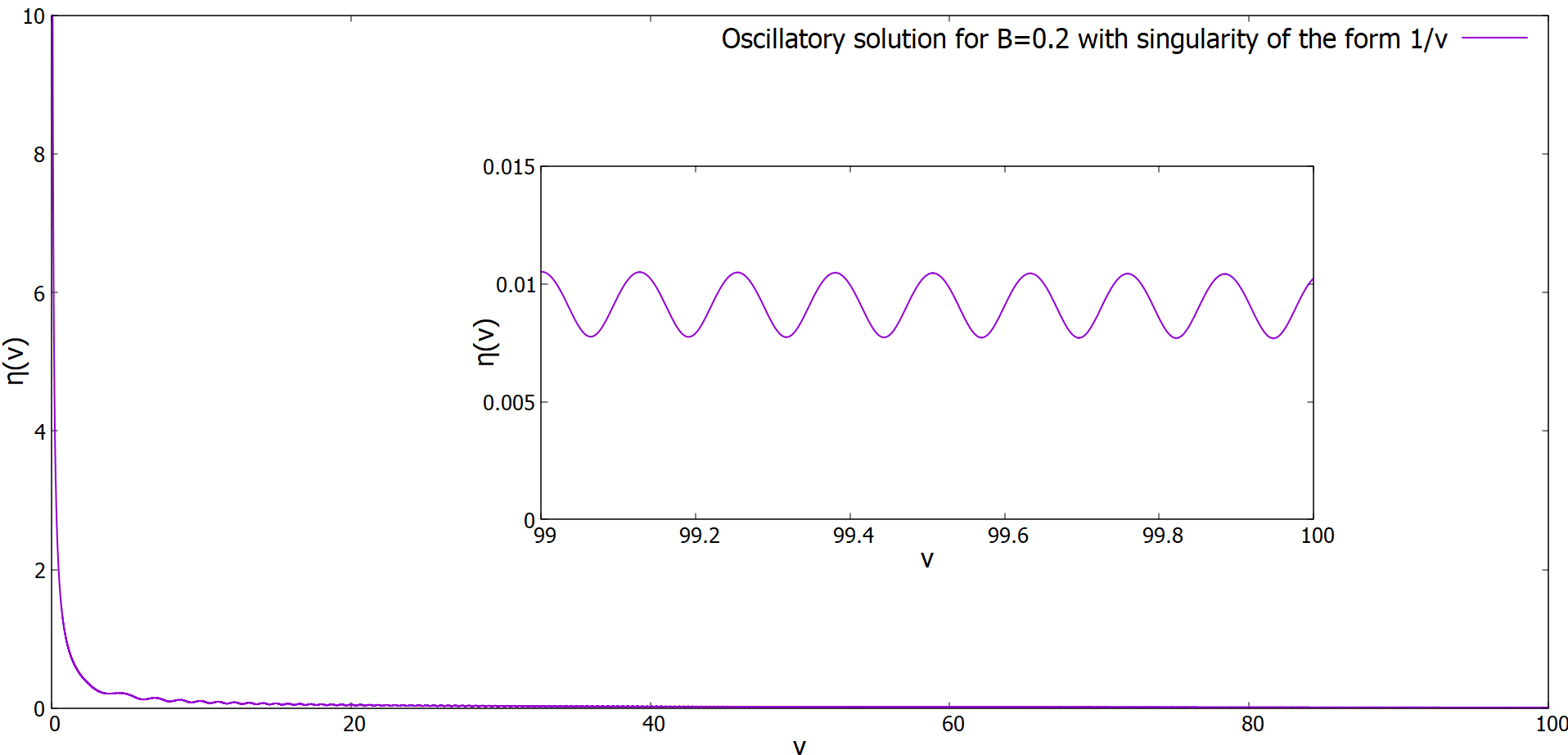}
        \caption{Sample solution} \label{fig:InvEta_v_B_0.2}
    \end{subfigure}
    \\[10pt]
    \begin{subfigure}{1\textwidth}
        \centering
        \includegraphics[width=\linewidth]{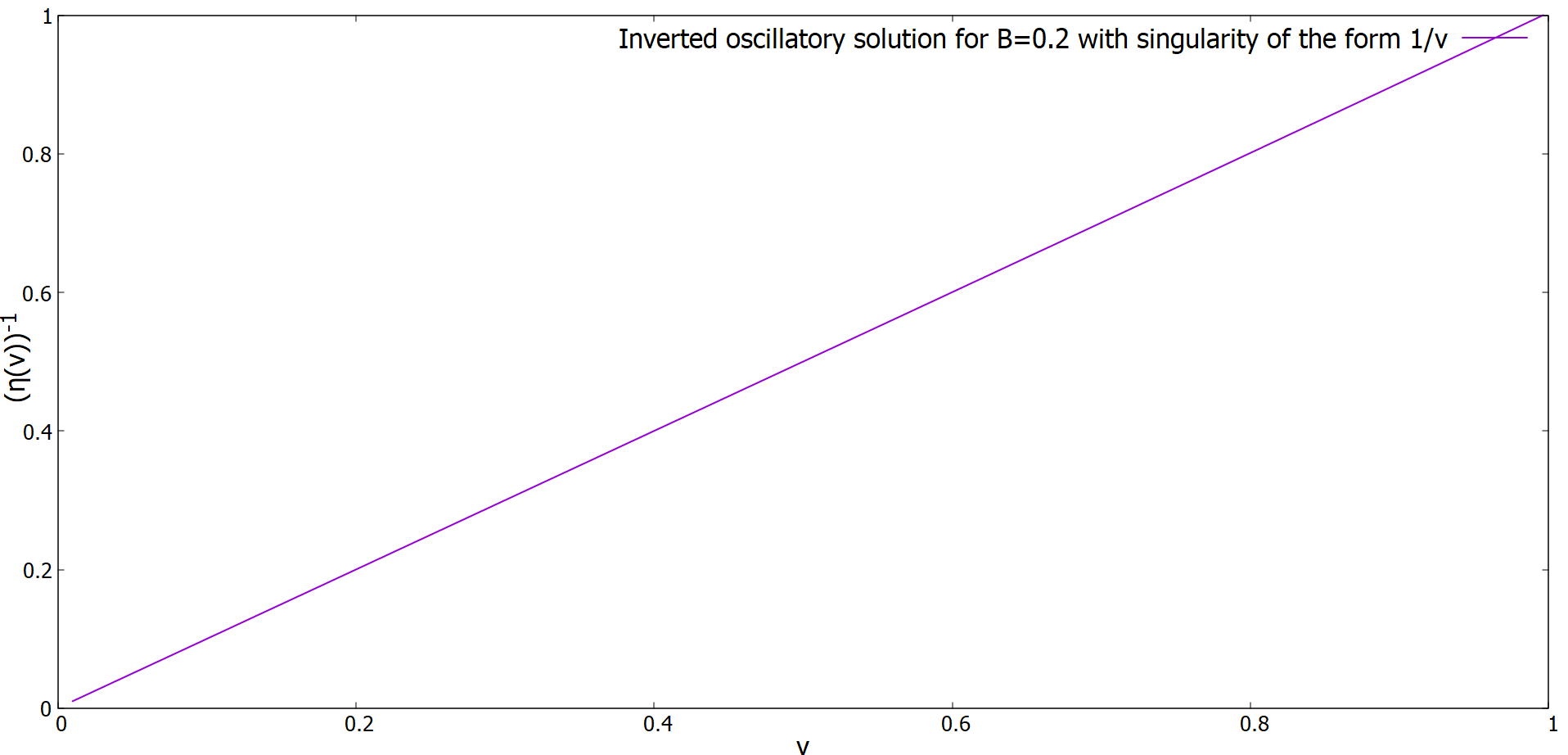}
        \caption{Inverted sample solution} \label{fig:InvEta_v_B_0.2_inverted}
    \end{subfigure}
    \vspace{5pt}
    \caption{Sample oscillatory solution with singularity of the form $\frac{1}{v}$ in 2D}
    \label{fig:InvEta_v}
\end{figure}

\begin{figure}[H]
    \centering
    \begin{subfigure}{1\textwidth}
        \centering
        \includegraphics[width=\linewidth]{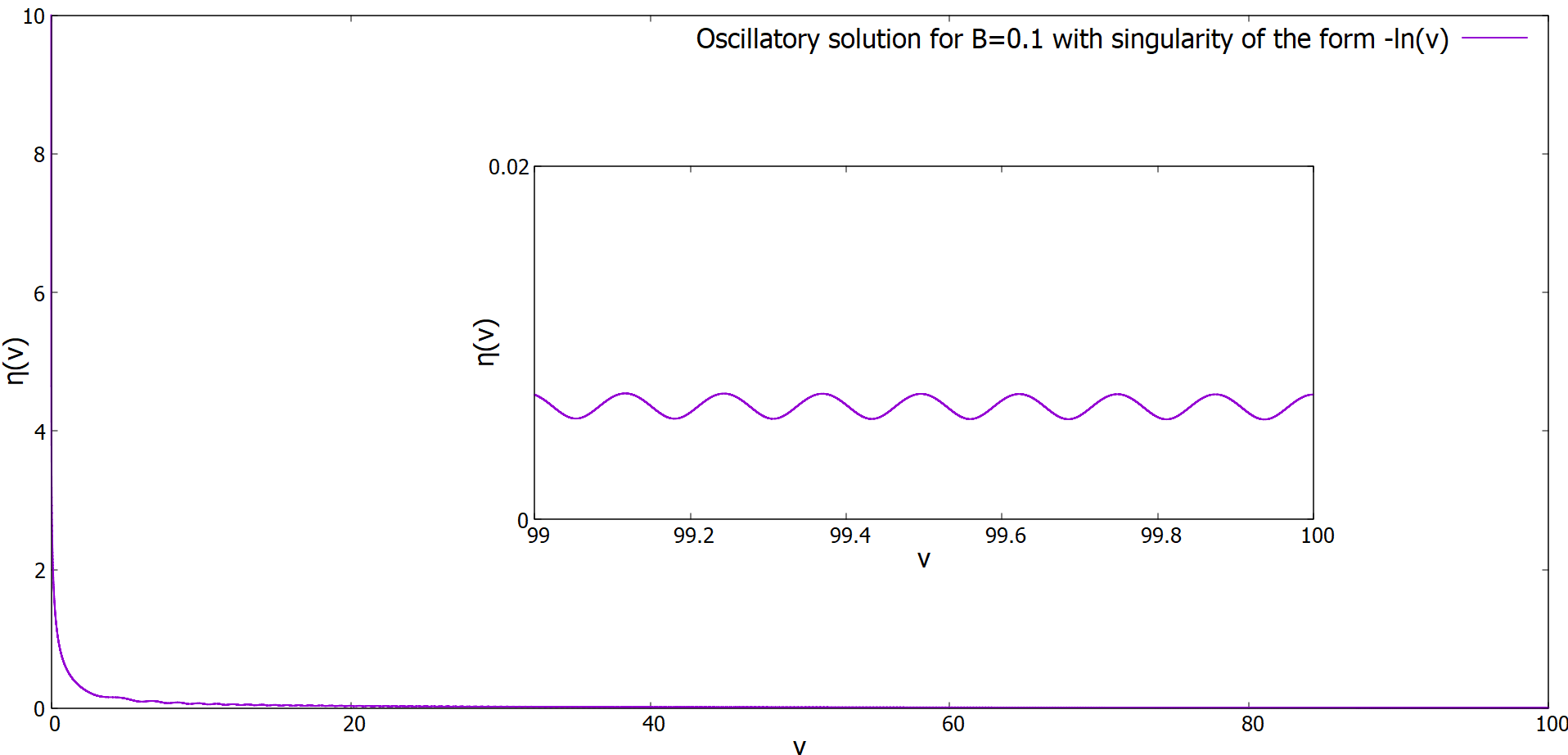}
        \caption{Sample solution} \label{fig:Logarithmic_solution_B_0.1}
    \end{subfigure}
    \\[10pt]
    \begin{subfigure}{1\textwidth}
        \centering
        \includegraphics[width=\linewidth]{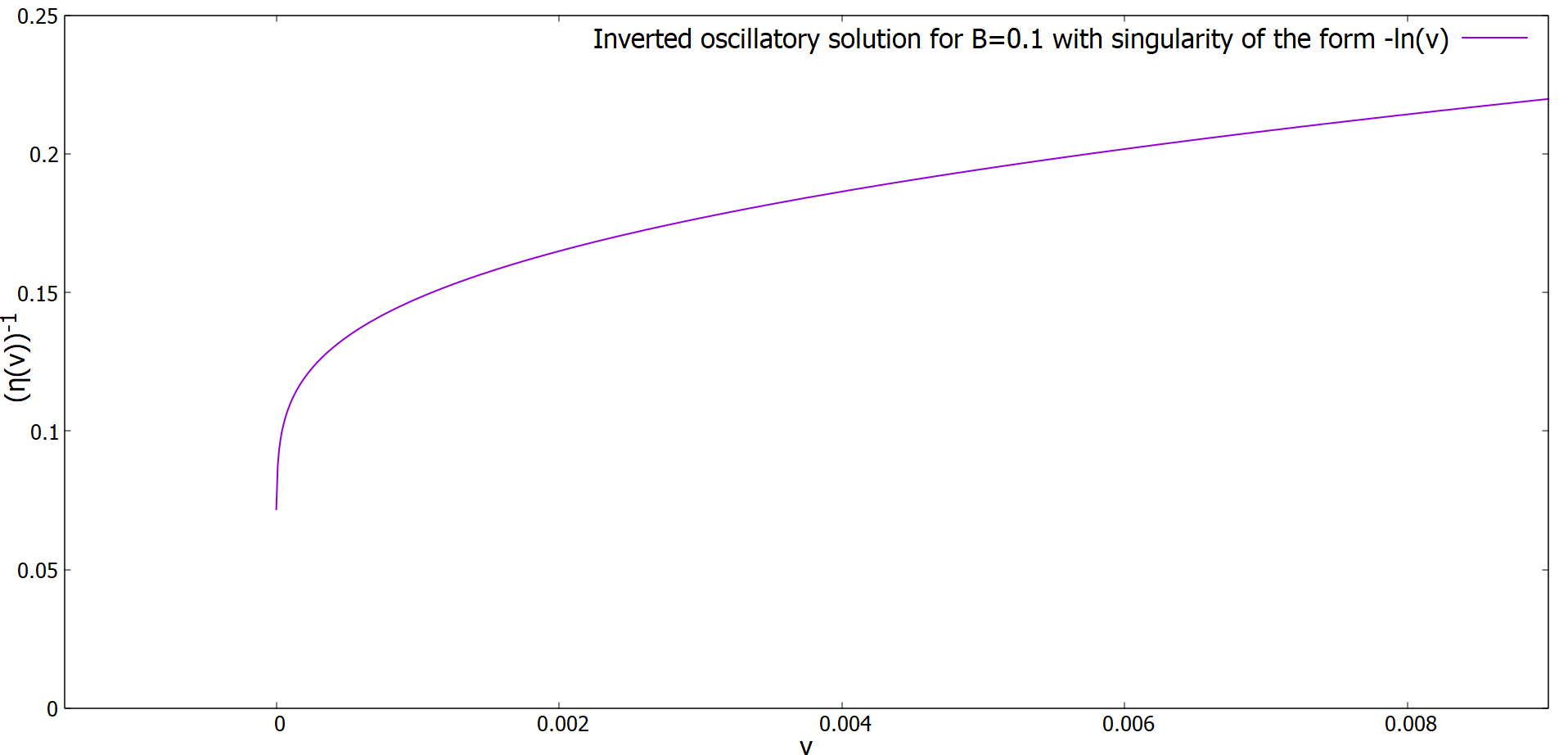}
        \caption{Inverted sample solution} \label{fig:Logarithmic_solution_B_0.1_inverted}
    \end{subfigure}
    \vspace{5pt}
    \caption{Sample oscillatory solution with logarithmic singularity at $v=0$ in 2D}
    \label{fig:Logarithmic_solution}
\end{figure}

The solutions (in 2D) discussed above are those with both transseries parameters being 0. However, we can also arbitrarily set the two parameters of the transseries subject to $C_2 = {C_1}^*$ as we saw earlier in \eqref{e:2DSelfSimilarODE_transseries_u}, which would be equivalent to setting the values of $\eta  \left(100 \right)$ and $\frac{d \eta  \left(v \right)}{d v}\Bigr|_{\substack{v=100}}$ that deviate from those given by equation~\eqref{e:2DSelfSimilarODE_series_v}. Then we can solve this initial value problem, the results of which are given in Figure \ref{fig:eta_0.01_etaPrime_-0.5}. These solutions clearly have oscillations consistent with non-perturbative asymptotic corrections of the form $\exp(\frac{\pm i v^{2}}{4})$ at large $v$. Note that the amplitude of the solution is of the form $\left|\chi  \left(R , T\right)\right| = \frac{1}{\sqrt{T}}\eta  \left(\frac{R}{\sqrt{T}}\right)$. This means that, when the singularity occurs at $v_{0} > 0$, it will move outward in the $R$ coordinate at the rate $\sim\sqrt{T}$. The only way to avoid this is to have a singularity that occurs at $v=0$. This behavior exists for non-oscillatory solutions we found for $B=\frac{1}{4}$ and $B=\frac{1}{8}$. To obtain oscillatory solutions with the singularity $\frac{1}{v}$, we solve the initial value problem for $\frac{1}{\eta  \left(v \right)} = \xi  \left(v \right)$ with $\xi  \left(0.01 \right) = 0.01$ and $\frac{d \xi  \left(v \right)}{d v}\Bigr|_{\substack{v=0.01}} = 1$ that give $\eta  \left(v \right)$ as shown in Figure \ref{fig:InvEta_v}. It seems that the oscillatory singular solutions with $\frac{1}{v}$ singularity exist only for $B \leq 0.5$. Similarly, if we picked the initial conditions from the series \eqref{e:2DSelfSimilarODE_series_v} at $v=100$ for $B=0.125$ for example, but set $B$ in ODE~\eqref{e:2DSelfSimilarODE} to some other small value (e.g., B=0.1, 0.2), we would get oscillatory solutions with logarithmic singularity as shown in Figure \ref{fig:Logarithmic_solution}.

\clearpage

\clearpage

\clearpage

\begin{figure}
    \centering
    \begin{subfigure}{1\textwidth}
        \centering
        \includegraphics[width=\linewidth]{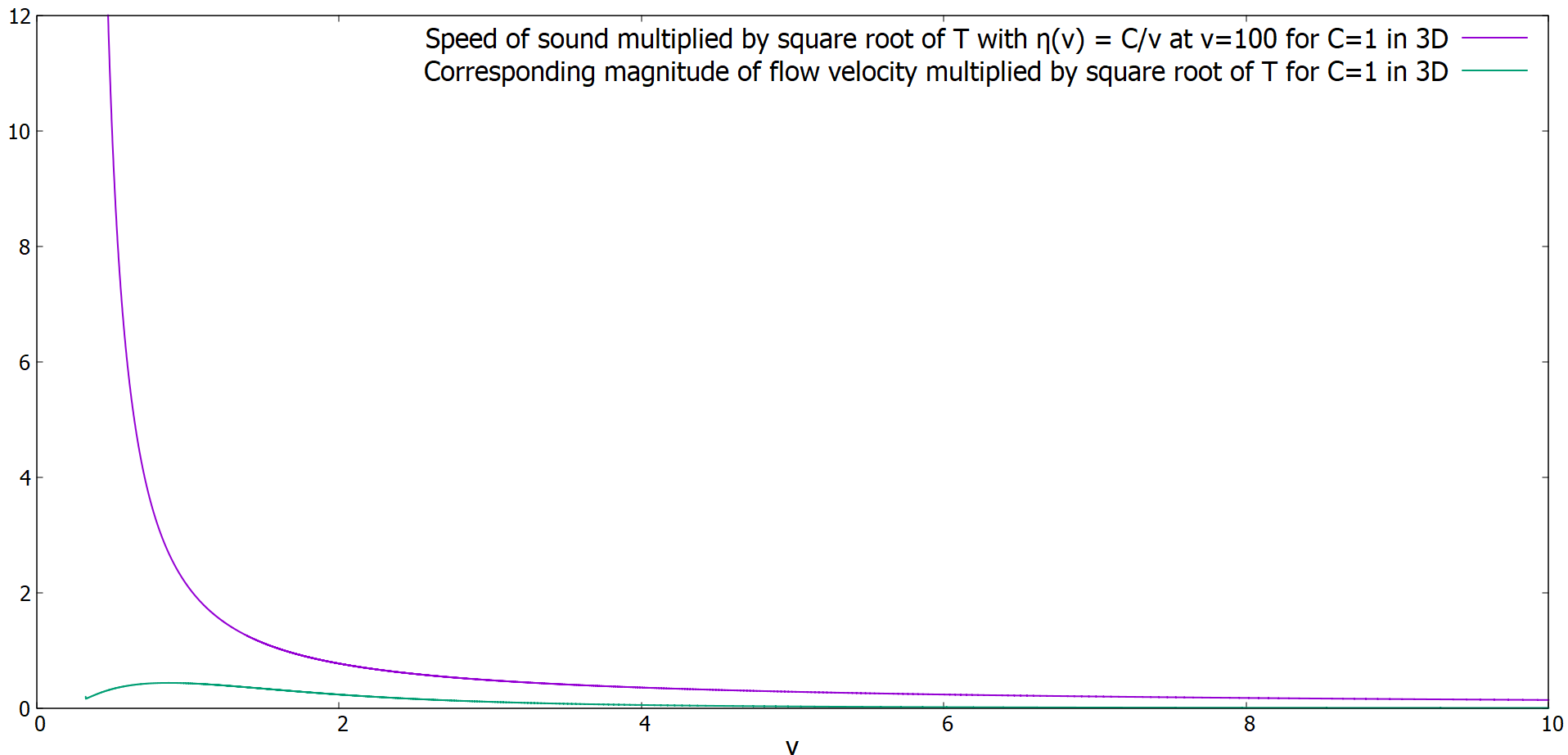}
        \caption{Solution with flow velocity approaching $0$ at infinity} \label{fig:C1_BH_3D}
    \end{subfigure}
    \\[10pt]
    \begin{subfigure}{1\textwidth}
        \centering
        \includegraphics[width=\linewidth]{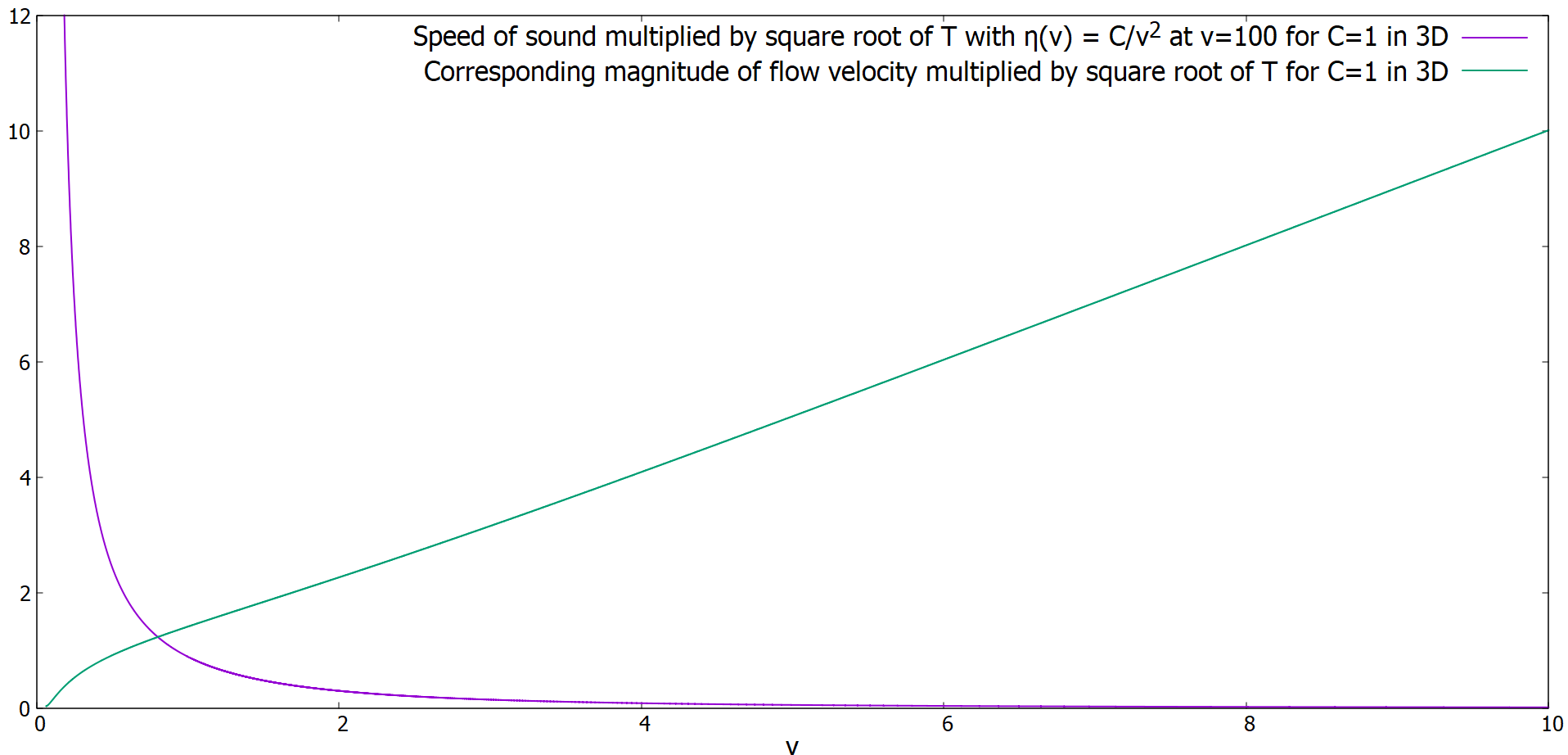}
        \caption{Solution with flow velocity approaching $v$ at infinity} \label{fig:C1_WH_3D}
    \end{subfigure}
    \vspace{5pt}
    \caption{Scaled sound speed and flow speed of BEC in coordinate $v$ for $C = 1$ in 3D}
    \label{fig:3D_SelfSimilar_numerical}
\end{figure}

\begin{figure}
    \centering
    \begin{subfigure}{1\textwidth}
        \centering
        \includegraphics[width=\linewidth]{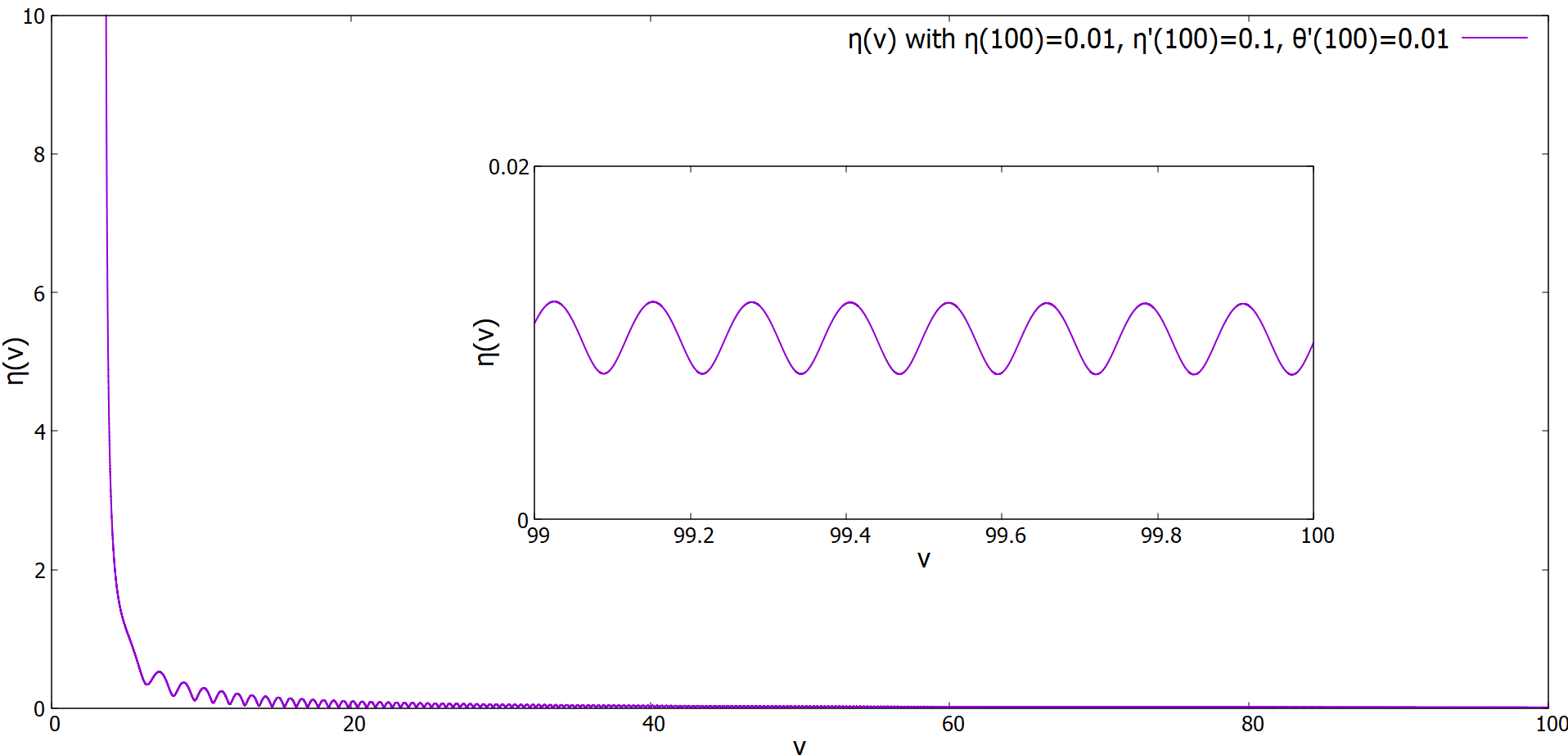}
        \caption{Sample solution} \label{fig:eta_0.01_etaPrime_0.1_thetaPrime_0.01}
    \end{subfigure}
    \\[10pt]
    \begin{subfigure}{1\textwidth}
        \centering
        \includegraphics[width=\linewidth]{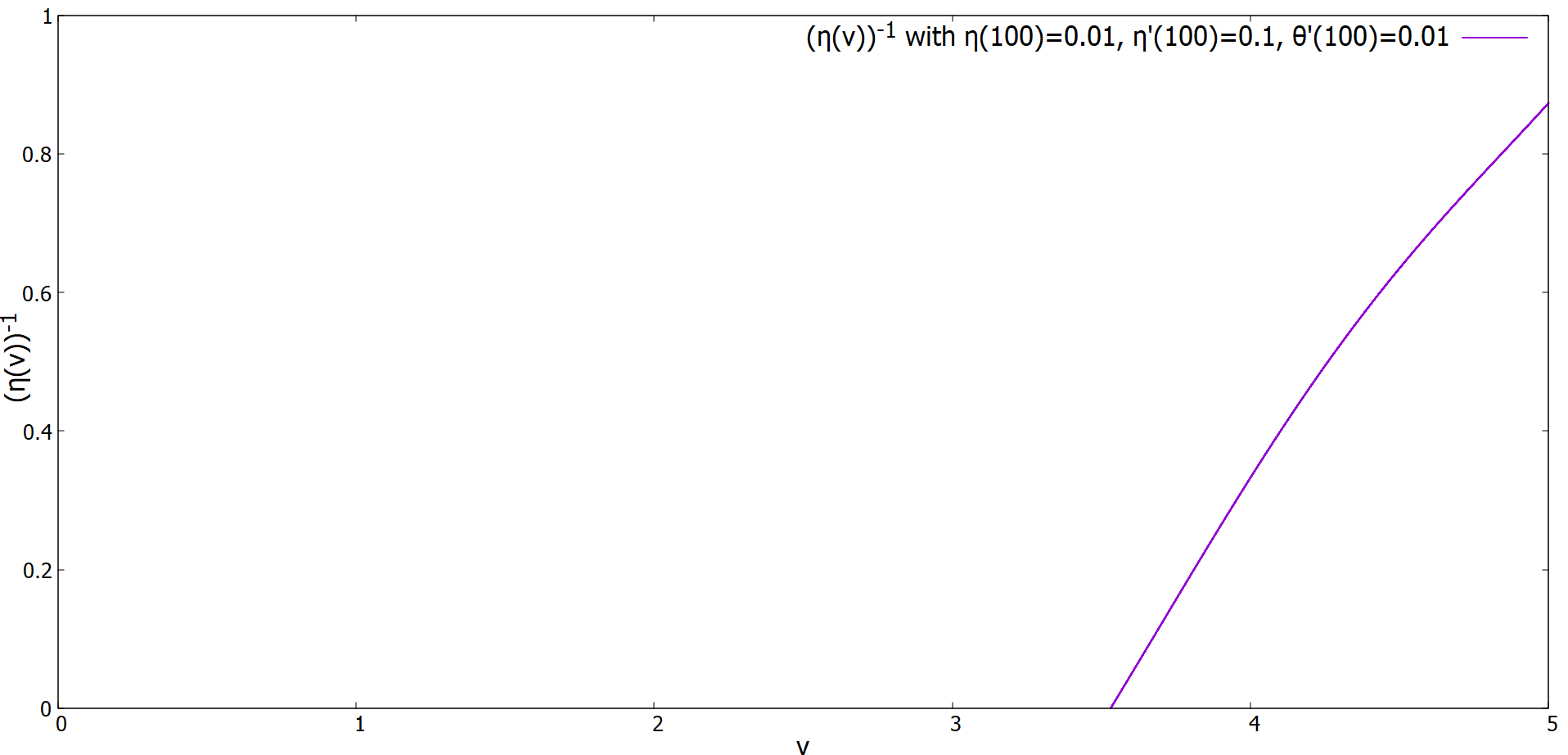}
        \caption{Inverted sample solution} \label{fig:eta_0.01_etaPrime_0.1_thetaPrime_0.01_inverted}
    \end{subfigure}
    \vspace{5pt}
    \caption{Sample oscillatory solution with singularity of the form $\frac{\sqrt{2}}{v-v_{0}}$ in 3D}
    \label{fig:Sample oscillatory solutions in 3D}
\end{figure}

\begin{figure}[H]
    \centering
    \begin{subfigure}{1\textwidth}
        \centering
        \includegraphics[width=\linewidth]{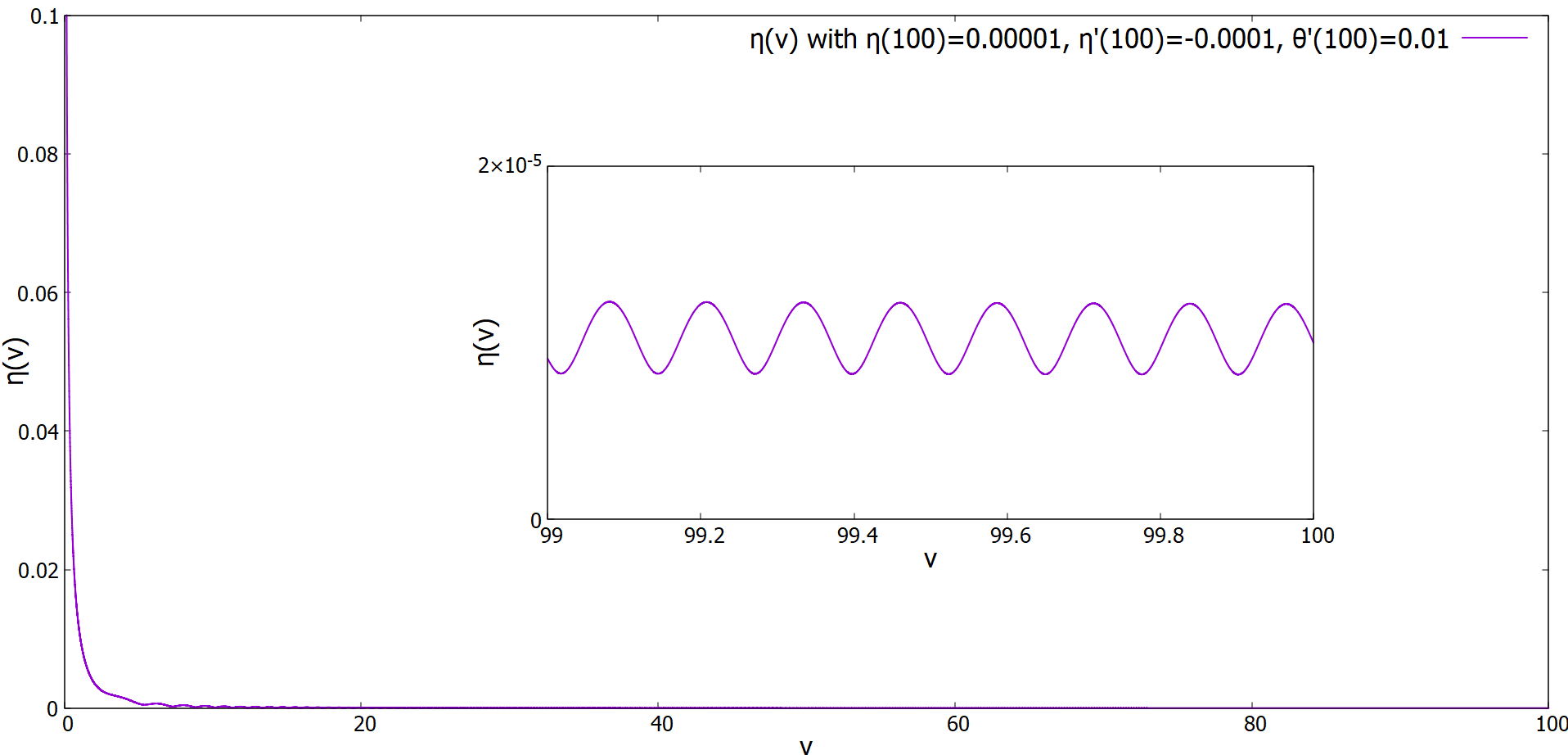}
        \caption{Sample solution} \label{fig:eta_eta_0.00001_etaPrime_-0.0001_thetaPrime_0.01}
    \end{subfigure}
    \\[10pt]
    \begin{subfigure}{1\textwidth}
        \centering
        \includegraphics[width=\linewidth]{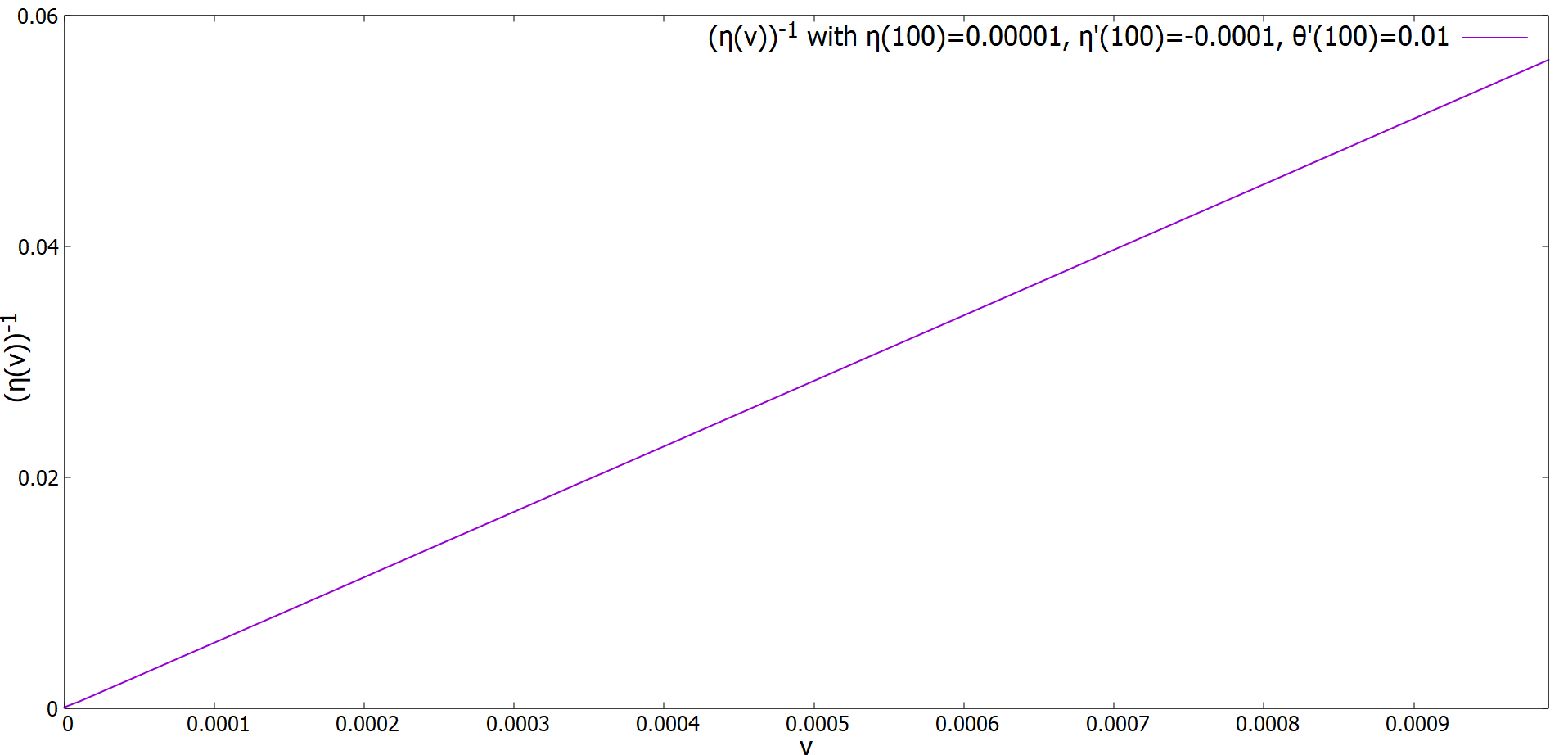}
        \caption{Inverted sample solution} \label{fig:eta_eta_0.00001_etaPrime_-0.0001_thetaPrime_0.01_inverted}
    \end{subfigure}
    \vspace{5pt}
    \caption{Sample (oscillatory) solution with singularity at $v\approx0$ in 3D}
    \label{fig:Sample (oscillatory) solutions with singularity at $v=0$ in 3D}
\end{figure}

\newpage
Now if we consider the 3D case, similar to the stationary case in \cite{vaidya2024stationaryacousticblackhole}, it can be shown that the governing differential equations \eqref{e:SelfSimilarReODE} and \eqref{e:SelfSimilarImODE} in 3D can also have singular solutions for $\eta(v)$. It can be shown that when the singularity occurs at some $v_0 > 0$, it is of the form $\frac{\sqrt{2}}{v-v_0}$ and should give $\frac{d \theta  \left(v \right)}{d v}\Bigr|_{\substack{v=v_{0}}} = \frac{v_0}{4}$. This is consistent with the result of the initial value problem. The initial value problem for a system of coupled ODEs \eqref{e:SelfSimilarReODE} and \eqref{e:SelfSimilarImODE} is solved with approximate initial conditions given by \eqref{e:asymptotic3D} starting from $v=100$ using RK4 \cite{Hairer2015}. Sample solutions (in 3D) with $C = 1$ are shown in Figure \ref{fig:3D_SelfSimilar_numerical}. Similar to the 2D case, even in 3D if we choose initial conditions (at $v=100$) deviating from the ones discussed in \eqref{e:asymptotic3D}, we see similar oscillatory behavior that appears to have oscillations of the form $\exp(\frac{\pm i v^{2}}{4})$ at large $v$, as can be seen in Figure \ref{fig:Sample oscillatory solutions in 3D}. For the same reason as 2D, we also find (oscillatory) singular solutions in 3D with singularity at $v\approx0$, by trial and error (Figure \ref{fig:Sample (oscillatory) solutions with singularity at $v=0$ in 3D}).

\clearpage

\clearpage

\clearpage

\section{Fluctuations in the Self-Similar background solution}
We can also study the dynamics of fluctuations in the background phase and amplitude (Self-Similar) solutions by writing the fluctuations as follows
\begin{align} \label{e:den phase pert SelfSimilar}
\chi \left(\Vec{\mathbf{R}}, T \right) & = \sqrt{n\left(\Vec{\mathbf{R}}, T \right)} \exp(i \theta\left(\Vec{\mathbf{R}}, T \right)) \nonumber \\ 
& = \sqrt{n_{0} \! \left(T, R \right) + n_{1} \! \left(\Vec{\mathbf{R}}, T \right)} \exp{(i (\theta_{0} \! \left(T, R \right) + \theta_{1} \! \left(\Vec{\mathbf{R}}, T \right)))} \nonumber \\ 
& = \sqrt{n_{0} \! \left(T, R \right)} \exp(i \theta_{0} \! \left(T, R \right)) \left(1 + N_{1} \! \left(\Vec{\mathbf{R}}, T \right) + i \theta_{1} \! \left(\Vec{\mathbf{R}}, T \right)\right) 
\end{align}
where $N_{1} \! \left(\Vec{\mathbf{R}}, T \right) = \frac{n_{1} \! \left(\Vec{\mathbf{R}}, T \right)}{2 n_{0} \! \left(T, R \right)}$. Here, $n_{0} \! \left(T, R \right) = {(\rho_{0} \! \left(T, R \right))}^2 = {\left(\frac{1}{\sqrt{T}} \eta \left(\frac{R}{\sqrt{T}} \right) \right)}^2$ and $\theta_{0} \! \left(T, R \right) = \theta_{0} \! \left(\frac{R}{\sqrt{T}} \right)$ are the background density and phase respectively. Additionally, $\Vec{\mathbf{R}}$ is $(R, \varphi)$ in 2D and $(R, \theta, \varphi)$ in 3D.

Substituting \eqref{e:den phase pert SelfSimilar} into equation \eqref{e:SelfSimilarPDE}, we get the real and imaginary parts of the linearized equations as follows
\begin{equation} \label{e:SS_density_pert}
\left({\partial}_T \pm \lvert \Vec{\mathbf{V}}  \left(R, T \right) \rvert {\partial}_R \right) N_1 \left(\Vec{\mathbf{R}}, T \right) = - \frac{1}{ {C \left(T, R \right)}^2} {\nabla}_R \cdot ({C \left(T, R \right)}^2 {\nabla}_R {\theta}_1 \left(\Vec{\mathbf{R}}, T \right))
\end{equation}

\begin{flalign} \label{e:SS_phase_pert}
\centering
\left({\partial}_T \pm \lvert \Vec{\mathbf{V}}  \left(R, T \right) \rvert {\partial}_R \right) {\theta}_1 \left(\Vec{\mathbf{R}}, T \right) & = \frac{1}{ {C \left(T, R \right)}^2} {\nabla}_R \cdot ({C \left(T, R \right)}^2 {\nabla}_R N_1 \left(\Vec{\mathbf{R}}, T \right)) \nonumber \\ & - {C \left(T, R \right)}^2 N_1 \left(\Vec{\mathbf{R}}, T \right)
\end{flalign}
where, $N_1 \left(\Vec{\mathbf{R}}, T \right) = \frac{n_{1} \! \left(\Vec{\mathbf{R}}, T \right)}{2 n_{0} \! \left(T, R \right)} = \frac{n_{1} \! \left(\Vec{\mathbf{R}}, T \right)}{2 \left(\rho_{0} \! \left(T, R \right) \right)^2} = \frac{n_{1} \! \left(\Vec{\mathbf{R}}, T \right)}{\left(C \left(T, R \right) \right)^2}$ and $\rho_{0} \! \left(T, R \right) = \frac{1}{\sqrt{T}} \eta \left(\frac{R}{\sqrt{T}} \right)$. Furthermore, it is easy to show under the change of coordinate $v = \frac{R}{\sqrt{T}}$ that the equations \eqref{e:SS_density_pert} and \eqref{e:SS_phase_pert} also exhibit self-similarity, and hence the phase fluctuations and density fluctuations are also self-similar and thus are static sinusoids in the $v = \frac{R}{\sqrt{T}}$ coordinate.

\newpage
Under the hydrodynamic approximation ${\nabla}_R \cdot ({C \left(T, R \right)}^2 {\nabla}_R N_1 \left(\Vec{\mathbf{R}}, T \right)) \approx 0$ (see \cite{CarlosBarceló_2001} and \cite{Visser2002}), equations \eqref{e:SS_density_pert} and \eqref{e:SS_phase_pert} can be combined into a wave equation that looks like a massless scalar field ${\theta}_1 \left(\Vec{\mathbf{R}}, T \right)$ in a (analog) space-time background. Corresponding analog space-time metric (see \cite{PhysRevA.63.023611}) in 2D turns out to be
\begin{equation} \label{e:acoustic metric self similar 2D}
g_{\mu \nu} \sim {\left(C \left(T, R \right) \right)}^4
\left[\begin{array}{ccc}
-\left(1-\frac{{\left(\lvert \Vec{\mathbf{V}} \left(T, R \right) \rvert \right)}^2}{{\left(C \left(T, R \right) \right)}^2}\right) & \mp\frac{\lvert \Vec{\mathbf{V}} \left(T, R \right) \rvert}{{\left(C \left(T, R \right) \right)}^2} & 0 
\\
 \mp\frac{\lvert \Vec{\mathbf{V}} \left(T, R \right) \rvert}{{\left(C \left(T, R \right) \right)}^2} & \frac{1}{{\left(C \left(T, R \right) \right)}^2} & 0 
\\
 0 & 0 & \frac{R^2}{{\left(C \left(T, R \right) \right)}^2} 
\end{array}\right]
\end{equation}
and analog space-time metric in 3D turns out to be
\begin{equation} \label{e:acoustic metric self similar 3D}
g_{\mu \nu} \sim {{\left(C \left(T, R \right) \right)}^3}
\left[\begin{array}{cccc}
-\left(1-\frac{{\left(\lvert \Vec{\mathbf{V}} \left(T, R \right) \rvert \right)}^2}{{\left(C \left(T, R \right) \right)}^2}\right) & \mp\frac{\lvert \Vec{\mathbf{V}} \left(T, R \right) \rvert}{{\left(C \left(T, R \right) \right)}^2} & 0 & 0 
\\
 \mp\frac{\lvert \Vec{\mathbf{V}} \left(T, R \right) \rvert}{{\left(C \left(T, R \right) \right)}^2} & \frac{1}{{\left(C \left(T, R \right) \right)}^2} & 0 & 0  
\\
 0 & 0 & \frac{R^2 {\sin}^{2}(\theta)}{{\left(C \left(T, R \right) \right)}^2} & 0
\\
0 & 0 & 0 & \frac{R^2}{{\left(C \left(T, R \right) \right)}^2}
\end{array}\right]
\end{equation}
where, ``$-$'' sign for outward flow and ``$+$'' sign for inward flow. Note that \eqref{e:acoustic metric self similar 2D} and \eqref{e:acoustic metric self similar 3D} resemble the Schwarzschild metric in Painlev\'e-Gullstrand coordinates (see \cite{HamiltonAndrewJ.S.2008Trmo}). We consider ``$-$'' sign in the acoustic metric tensors above because these are the solutions with crossover between local speed of sound and flow speed indicating an acoustic (white hole) horizon.

\section{Self-Similar behavior}
We see that self-similar non-oscillatory solutions seem to have outward flow velocity indicating a sonic analog of a white hole when there is a crossing between flow speed and sound speed. The same is the case with some self-similar oscillatory solutions\footnote[7]{They correspond to the transseries with at least one of the two real transseries parameters being nonzero as discussed earlier.}. It is easy to see that the functions $\eta(v)$ and $\frac{d}{dv}\left(\theta(v)\right)$ appear to be fixed in coordinate $v$. This indicates that all functions (in the solution) in $v$, when mapped back to the coordinate $R$, move radially outwards (especially oscillatory ones in the form of sinusoidal waves) with time at the rate $\propto \sqrt{T}$. Furthermore, from $\frac{\sqrt{2}}{\sqrt{T}}\eta(v)$ and $\frac{2}{\sqrt{T}}\frac{d}{dv}\left(\theta(v)\right)$, all amplitudes, sound speed, and flow velocity clearly also decay with time at the rate $\propto\frac{1}{\sqrt{T}}$. Furthermore, since the flow is always outgoing in 2D and 3D, when there is a crossing between the approximate local speed of sound and the magnitude of the flow velocity, there is a white hole horizon that remains at a constant $v$ defined by ${\left(2\frac{d}{d v}  \left(\theta \left(v \right) \right) \hat{\mathbf{R}}\right)}^{2} - {\left(\sqrt{2} \eta  \left(v \right)\right)}^{2} = 0$ from \eqref{e:acoustic metric self similar 2D}, \eqref{e:acoustic metric self similar 3D}, \eqref{e:scaled_sound_ss}, and \eqref{e:scaled_vel_ss}. Therefore, the radius of the horizon grows as $R_{h} \propto \sqrt{T}$.

\section{Conclusions}
In this article, we studied Self-Similar (in radial coordinate and time) configurations of non-relativistic Bose-Einstein condensate (BEC) described by the Gross-Pitaevskii Equation (GPE). The (singular) self-similar solutions in this model in 2D (with circular symmetry) and 3D (with spherical symmetry) have outward flow velocity, and it crosses the scaled local speed of sound, clearly indicating the existence of a sonic analog of a white hole. Particularly in 3D, this crossing occurs in the case of the solutions with flow velocity growing radially outward. Since the flow remains supersonic everywhere outside the horizon and keeps getting faster with radial distance, this model is more relevant for analog models in a cosmological context (see \cite{PhysRevA.69.033602}, \cite{PhysRevLett.91.240407} for example) in the non-relativistic limit. Furthermore, the white hole horizon grows at the rate $R_{h}\propto\sqrt{T}$ as we saw. Also, the oscillatory solutions appear like sinusoidal waves moving outward at the rate of $\sqrt{T}$ while simultaneously being suppressed by the factor of $\frac{1}{\sqrt{T}}$. We also find that the linearized equations governing fluctuations  as well as the acoustic metric tensors also exhibit this self-similarity.

In 2D, we also check the Transseries\footnote[8]{With both (real) transseries parameters set to $0$ for the reasons explained earlier.} as $v=\frac{R}{\sqrt{T}}\to\infty$ and demonstrate how its Laplace-Borel resummation matches well with the numerical solutions obtained from the Runge-Kutta method of order four (RK4). We show how Laplace-Borel resummation can be used to extract information about the solutions even at smaller $v=\frac{R}{\sqrt{T}}$ values from the limited amount of information available at $v=\frac{R}{\sqrt{T}}\to\infty$.

\section{Acknowledgments}

I am grateful to the DOE for support under the Fermilab Quantum Consortium. I would like to thank my PhD advisor, Dr. Luis M. Kruczenski, for his guidance and support for this project. I wish to acknowledge valuable discussions with Akilesh Venkatesh and Kunaal Joshi on numerical techniques. I also wish to acknowledge valuable computational resources provided by RCAC (Rosen Center For Advanced Computing) at Purdue University.

\printbibliography

\end{document}